\documentstyle[12pt]{report}
\let\tilde=\widetilde
\def\e{{\rm e}}
\def\i{{\rm i}}
\def\etal{{\sl et al.}}
\def\etc{{\sl etc.}}
\def\tr{{\rm tr}}
\def\qdet{\hbox{q-det}}
\def\det{\mathop{\rm det}\nolimits}

\def\one#1{#1^{\raise4pt\hbox{$\scriptstyle\!\!\!\!1$}}\,{}}
\def\two#1{#1^{\raise4pt\hbox{$\scriptstyle\!\!\!\!2$}}\,{}}
\def\three#1{#1^{\raise4pt\hbox{$\scriptstyle\!\!\!\!3$}}\,{}}
\def\vac{\left|0\right>}
\def\phi{\varphi}

\def\a{\alpha}
\def\d{\delta}
\def\D{\Delta}
\def\l{\lambda}
\def\L{\Lambda}
\def\s{\sigma}
\def\beq{\begin{equation}}
\def\eeq{\end{equation}}
\def\bds{\begin{description}}
\def\eds{\end{description}}
\def\half{\frac{1}{2}}
\def\halfe{\frac{\eta}{2}}
\newtheorem{cond}{Condition}
\newtheorem{th}{Theorem}[chapter]

\def\TR{${\cal T}_R$}
\def\XD{${\cal X}_\Delta$}
\def\x{{\bf x}}
\def\X{{\bf X}}
\def\y{{\bf y}}
\def\Y{{\bf Y}}

\def\Z{{\bf Z}}
\def\b{{\bf b}}
\def\B{{\bf B}}
\def\R{{\cal R}}
\def\K{{\cal K}}
\def\BB{{\cal B}}
\def\PP{{\cal P}}
\def\tt{\widetilde{t}}
\def\DD{\widetilde{\Delta}}
\newcommand{\bb}{\hat{b}}
\newcommand{\C}{{\bf C}}
\newcommand{\Fun}{{\rm Fun}}
\newcommand{\Sym}{{\rm Sym}}
\newcommand{\sov}{separation of variables}
\begin{document}
\begin{flushright}
{\sf HU-TFT-91-51,} 7 October 1991 \\
{\sf hep-th/9211111 }
\end{flushright}
\vskip 1.5cm
\begin{center}
{\LARGE\bf Quantum Inverse Scattering Method. Selected Topics}
\end{center}
\vskip 1.5cm
\noindent Four lectures given at Nankai Institute of Mathematics,
  Tianjin, China, 5--13 April 1991
\vskip 1.5cm
\begin{center}
{\Large\bf E K Sklyanin}
\end{center}
\vskip 1cm
\footnote[1]{The work supported by the Academy of Finland}
\footnote[2]{Permanent address:
St.Petersburg Branch of Steklov Mathematical Institute, Fontanka 27,
St.Petersburg 191011, USSR}
Research Institute for Theoretical Physics, University of Helsinki,
Siltavuorenpenger 20C, 00170 Helsinki 17, Finland
\vskip 1cm
{\bf Abstract.} The lectures present an elementary introduction into the
quantum integrable models aimed for mathematical physicists and
mathematicians. The stress is made on the algebraic aspects of the theory
and the problem of determining the spectrum of quantum integrals of motion.
The XXX magnetic chain is used as the basic example. Two lectures are
devoted to a detailed exposition of the Functional Bethe Ansatz --- a new
technique alternative to the Algebraic Bethe Ansatz --- and its relation to
the separation of variable method. A possibility to extend FBA to the
$SL(3)$ is discussed.
\vskip 1.5cm
{\sl Published in ``Quantum Group and Quantum Integrable Systems'' (Nankai
Lectures in Mathematical Physics), ed.\ Mo-Lin Ge, Singapore:
World Scientific, 1992, pp.\ 63--97.}

\newpage
\baselineskip=6mm
\setcounter{chapter}{1}
\begin{center}
    \large\bf Lecture 1
\end{center}
It is great pleasure for me to visit China and to take part in this
school. I would like to thank the Nankai Institute of Mathematics
and Professor Ge Mo-Lin for hospitality.

My lectures will be devoted to the Quantum Inverse Scattering
Method (QISM). The QISM is a direction in the theory of
quantum integrable systems which starts its history from
the summer 1978 when 3 groups: in Leningrad, USSR (Faddeev \etal),
Fermilab, USA (Thacker, Creamer, Wilkinson) and in Freiburg, Germany
(Honerkamp \etal), studying the quantum nonlinear Schr\"{o}dinger
equation, had found some striking and puzzling connections between
the famous Bethe Ansatz and  no less famous Classical Inverse
Scattering Method (CISM).

In other words, QISM has arisen as a result of a synthesis of two
traditions in the theory of integrable systems which, up to 1978,
were developing quite independently. The first tradition originates
from the famous paper by Hans Bethe (1931) from which the Bethe
Ansatz took its name. It was developed in the works of L.Hulthen,
E.Lieb, C.N.Yang, C.P.Yang, R.J.Baxter and many other scholars
devoted mainly to exactly soluble models of lattice statistical
mechanics and of quantum mechanics. The second, more young,
tradition originates from the
paper of Gardner, Green, Kruskal and Miura on the KdV equation which
gave rise to CISM
(Lax, Ablowitz, Kaup, Newell, Zakharov, Shabat, Faddeev,... ).
Still, the roots
of this tradition can be traced even deeper into the mathematics
of 19th century (Liouville, Jacobi, Kowalewsky,...).

One should mention also two more directions which contributed
considerably to QISM. These are the Factorizable $S$-matrices theory
due to A. B. Zamolodchikov and Al. B. Zamolodchikov and the various
group-theoretical approaches
(Adler, Kostant, Olshanetsky, Perelomov, Semenov--Tian--Shansky,
Reyman \etal).

I am not going to present here the complete history of QISM. However, I
would like to mention the main successes of QISM: exact quantization of
the sine-Gordon equation (Faddeev, Takhtajan) and calculation of correlators
for various quantum integrable models (V. E. Korepin, N. M. Bogoliubov,
A. G. Izergin, F. A. Smirnov).
Witnessing the contemporary Quantum Groups boom I cannot help reminding you
that the whole QG business has arosen as a by-product of QISM, being now quite
independent discipline. Let me add also one more remark concerning the
development of QISM. There is a general trend in QISM which appears to
become stronger last years: shift of stress from applications and related
analytical questions to algebraic structures underlying the integrability.

In my lectures  I am going to give an elementary introduction to QISM
and to touch also some more special questions (Functional Bethe Ansatz).
Since I would like to concentrate on the mathematical methods involved
rather than applications of QISM, in the center of our attention there will be
the only problem: calculation of the spectrum of integrals of motion.
Various approaches will be illustrated on the sole example:
$XXX$ magnetic chain.

To begin with, let me discuss briefly the concept itself of integrability.
In the classical mechanics there is well-known definition of
integrability due to Liouville (1855). According to Liouville,
the classical Hamiltonian finite-dimensional system is called integrable if
it possesses a set of independent integrals of motion commuting
with respect to the Poisson bracket
$$ \{I_j,I_k\}=0. $$
The total number of the integrals of motion (including Hamiltonian)
should be the half of the dimension of the phase space.

The Liouville's theorem provides also a way of constructing the
action-angle variables for an integrable system in terms of the
multi-variable curvilinear integrals. Being quite good for
theoretical studies, this construction, unfortunately,
does not help to perform an effective integration of concrete models.
In practice, one needs usually to resort to more special techniques
such as CISM or some algebraic methods.

The situation with the definition of the quantum integrability is
even worse. The first thing which comes to mind is to mimic
the classical Liouville's definition, namely, to require the
existence of N ($=$ number of degrees of freedom)  commuting
operators
$$ [I_j,I_k]=0.             $$
%{ useful but not enough!}
Unfortunately, it seems to be very hard to develop this idea up to
a satisfactory level of rigor. The main obstacle is the difficulty
with correct definition of functional independence of integrals of
motion in the quantum case. It is certain that it is necessary to
restrict somehow the class of allowed functions but,
as far as I know, at the moment there is no consistent
theory of that kind. To sum up, I must confess that I don't know any
good definition of the quantum complete integrability.

Let me take during these lectures the most pragmatical point of view:
I shall call a quantum system ``integrable'' if it is possible
to calculate exactly some quantities of physical interest, such as
the common spectrum of commuting quantum integrals of motion
or some correlators (in these lectures I'll concentrate on the first
problem: the spectrum). The word ``exactly'' needs, of course, some comment.

%{It is not an exaggeration to say that the mathematical physics of 19th
% century developed...}

An excursion into the history of classical mechanics  and
mathematical physics shows that the concept of ``exact solubility''
has been changed during the course of time. The main trend was
the permanent extension of the class of functions used. First,
elementary functions, then integrals of them, then solutions to
certain second order differential equations, elliptic functions
\etc. The most general concept of exact solubility elaborated
in 19th century is, in my opinion, the concept of {\it separation
of variables}, that is, reduction of a multidimensional problem to
a series of one-dimensional ones. In my 3rd and 4th lectures
devoted to the ``Functional Bethe Ansatz'' I'll try to show you
that this old concept works quite effectively in the framework of
QISM.

%{Algebraic approach}
Now let us start the systematic introduction to QISM. The basic idea
of QISM, as it has become clear the last years, is purely algebraic.
Its roots can be traced up to the
very dawn of quantum mechanics. I mean the early matrix formulation
of quantum mechanics due to Heisenberg which led to a purely
algebraic treatment
of the quantum harmonic oscillator (Heisenberg, 1925)
and the hydrogen atom (Pauli, 1926).
The idea is to include the commutative
algebra of the quantum integrals of motion $\{I_n\}$ into some bigger
algebra ${\cal A}$. The space of the quantum states of the system in question
is considered then as a representation (usually, irreducible) of that
bigger algebra ${\cal A}$ whose elements produce,
roughly speaking, the transitions
between the eigenstates of the quantum conserved quantities $\{I_n\}$.
The common spectrum of $\{I_n\}$ can be found then by purely algebraic means.

In case of the harmonic oscillator there is only one integral of motion,
the number of particles operator $N$, and it can be included into the
Heisenberg Lie algebra ${\cal A}$ generated by $N$ and three extra generators:
$h$(central element), $a$, $a^\dagger$
$$ [h,a]=[h,a^\dagger]=[h,N]=0$$
$$ [a,a^\dagger]=h$$
$$ [N,a]=-a \qquad [N,a^\dagger]=a^\dagger $$
The procedure of finding the spectrum of $N$ using the creation/annihilation
operators $a$, $a^\dagger$ is well known.

In case of the hydrogen atom (Coulomb problem)
the components of the angular momentum vector
and the so-called Laplace vector
form an algebra whose factor over the relation
{\it Hamiltonian$=$const} is
the $so(4)$ Lie algebra. This remarkable fact allows to determine the spectrum
of the Hamiltonian  by purely algebraic methods.

In the above examples the bigger algebra ${\cal A}$ was,
up to minor reservations in the hydrogen atom case,
 a finite-dimensional
Lie algebra. The fundamental peculiarity of QISM consists in using
a new class of algebras to describe the dynamical symmetry of quantum
integrable systems. These algebras are neither finite-dimensional
nor Lie algebras.

The algebras used in QISM are described in terms of the generators
$T_{\alpha\beta}(u)$, $\alpha,\beta\in\{1\ldots d\}$ which can be
considered as the elements of the square $d\times d$ matrix $T(u)$
depending on the continuous parameter $u$ frequently called the
spectral parameter. The associative algebra \TR\ is generated
then by the quadratic relations

{\samepage
$$\sum_{\beta_1,\beta_2=1}^d
   R_{\alpha_1\alpha_2,\beta_1\beta_2}(u-v)
   T_{\beta_1\gamma_1}(u)T_{\beta_2\gamma_2}(v)$$
\nopagebreak
$$=\sum_{\beta_1,\beta_2=1}^d
 T_{\alpha_2\beta_2}(v)T_{\alpha_1\beta_1}(u)
 R_{\beta_1\beta_2,\gamma_1\gamma_2}(u-v)\quad \forall u,v$$}
where $R_{\alpha_1\alpha_2,\beta_1\beta_2}(u)$ is some given function
(``structure constants tensor''). Using the matrix notation
$$ \one T\equiv T\otimes{\rm id}\qquad
   \two T\equiv {\rm id}\otimes T$$
the former relation can be written down in the compact form
\beq
R(u-v)\one T(u)\two T(v)=\two T(v)\one T(u)R(u-v)
\label{eq:RTT}
\eeq

The ``structure constants'' $R$ are required to satisfy the
consistency condition (the well known nowadays Yang--Baxter equation)
$$\sum_{\beta_1,\beta_2,\beta_3=1}^d
  R_{\alpha_1\alpha_2,\beta_1\beta_2}(u)
  R_{\beta_1\alpha_3,\gamma_1\beta_3}(u+v)
  R_{\beta_2\beta_3,\gamma_2\gamma_3}(v)$$
$$=\sum_{\beta_1,\beta_2,\beta_3=1}^d
  R_{\alpha_2\alpha_3,\beta_2\beta_3}(v)
  R_{\alpha_1\beta_3,\beta_1\gamma_3}(u+v)
  R_{\beta_1\beta_2,\gamma_1\gamma_2}(u)\quad\forall u,v$$
or, briefly,
\beq
  R_{12}(u)R_{13}(u+v)R_{23}(v)=
  R_{23}(v)R_{13}(u+v)R_{12}(u)
\label{eq:YBE}
\eeq
(the notation is obvious).

It is easy to see that the matrix trace $t(u)$ of $T(u)$
$$t(u)=\tr T(u)\equiv\sum_{\alpha=1}^d T_{\alpha\alpha}(u)$$
forms a commutative family of operators
$$[t(u),t(v)]=0\qquad\forall u,v$$
which can be thus considered as integrals of motion of some
quantum integrable system. Generally speaking, there can be other
independent integrals of motion but in the case $d=2$, to which
I'll restrict my attention in these lectures, $t(u)$ turns out
to be the maximal commutative subalgebra of \TR.

So, given a solution $R(u)$ of the Yang--Baxter equation
one can define the quadratic algebra \TR. Given a representation
of the algebra \TR\  one obtains a quantum integrable system whose
quantum space is the representation space of \TR\  and the
commutative integrals of motion are $t(u)$. The main problem of QISM is to
find their common spectrum and, possibly, correlators of some
physically interesting operators.

{\samepage
The main steps of QISM can be summarized as follows:
\begin{enumerate}
\item Take an R matrix.
\item Take a representation of \TR.
\item Find spectrum of $t(u)$.
\item Find correlators.
\end{enumerate}}

The first step implies solving the Yang--Baxter equation (YBE).
Many particular solutions has been found by trial-and-error method.
As for for the general theory of YBE, the major contribution has been made
by V. Drinfeld who gave an axiomatics of QISM based on the concept of
{\it Hopf algebras}. In Drinfeld's axiomatics the steps 1 and 2 of our
scheme are intertwinned inseparably. In the base of his theory lies the
concept of the {\it quasitriangular Hopf algebra} whose representations produce
particular $R$-matrices. Drinfeld has constructed also an important family
of quasitriangular Hopf algebras called {\it Yangians} ${\cal Y}[{\cal G}]$
and parametrized by a simple Lie algebra ${\cal G}$.

I don't intend to go further into details of Drinfeld's theory.
For these lectures I have chosen another approach supposing that
a solution to the YBE is given from the very beginning. This standpoint
lies  closer to the original form of QISM as it
appears in the works of Leningrad group and is more convenient for
applications and for discussing the main topic of these lectures ---
the spectrum of $t(u)$ (Step 3).

{}From this very moment I'll consider only the case $d=2$ that is
$T(u)$ being a $2\times2$ matrix
$$ T(u)=\left(\begin{array}{cc}
               A(u) & B(u)\\
               C(u) & D(u)\end{array}\right)          $$
Moreover, I'll restrict my attention to the simplest solution
$R(u)$ of the YBE, $R$--matrix of the XXX--magnet, corresponding to the
Yangian ${\cal Y}[sl(2)]$.
The solution is expressed in terms of the permutation operator
${\cal P}$ in the tensor product ${\bf C}^2\otimes{\bf C}^2$
$$ {\cal P}x\otimes y =y\otimes x\qquad \forall x,y\in{\bf C}^2$$
and reads as
\beq
  R(u)=u+\eta{\cal P}=\left(\begin{array}{cccc}
     a&&&\\
     &b&c&\\
     &c&b&\\
     &&&a\end{array}\right)\qquad
  \begin{array}{rcl}
  a&=&u+\eta\\
  b&=&u\\
  c&=&\eta\end{array}
\label{eq:RXXX}
\eeq
in the natural basis
$\{e_1\otimes e_1,e_1\otimes e_2,e_2\otimes e_1,e_2\otimes e_2\}$
in ${\bf C}^2\otimes{\bf C}^2$.

Let us discuss now the second step of our scheme: finding
representations of  \TR\  for given $R(u)$. The algebra
\TR\  possesses an important property which is
called {\it comultiplication}. Let $T_1(u)$ and $T_2(u)$ be two
representations of \TR\  in the spaces $V_1$ and $V_2$
respectively. Then the matrix
$$ T(u)=T_1(u)T_2(u)\qquad T_{\alpha\gamma}(u)\equiv
    \sum_{\beta=1}^d T_{1,\alpha\beta}(u)T_{2,\beta\gamma}(u)$$
is also a representation of \TR\  in the space $V_1\otimes V_2$
called {\it tensor product} of representations $T_1(u)$ and $T_2(u)$.
The possibility to multiply the representations of \TR\
provides immediately an opportunity to construct infinitely many
representations given a set of particular representations.
Usually such basic representations are chosen to have some simple
dependence (say, polynomial) on the spectral parameter $u$.
Traditionally, in QISM such elementary representations are called
{\it L--operators} and their product
$$T(u)=L_N(u)\ldots L_2(u)L_1(u)   $$
resp.  {\it monodromy matrix}. Every new representation of this type
gives rise to a new quantum integrable system.
Putting aside the question of
completeness of such representations, I would like to remark
that this family of representations serves perfectly all the
models important for applications.

Let us turn now to our basic example, the algebra \TR\
corresponding to the XXX R--matrix described above.
This algebra possesses the remarkable central element
(Casimir operator) called {\it quantum determinant}
\beq
  \begin{array}{rcl}
  \D(u)\equiv\qdet T(u)&=&{\rm tr}_{12}\frac{1-{\cal P}}{2}
     \one T(u-\halfe)\two T(u+\halfe)\\
&=&{\rm tr}_{12}
     \two T(u+\halfe)\one T(u-\halfe)\frac{1-{\cal P}}{2}\\
&=&D(u-\halfe)A(u+\halfe)-B(u-\halfe)C(u+\halfe)\\
&=&A(u-\halfe)D(u+\halfe)-C(u-\halfe)B(u+\halfe)\\
&=&A(u+\halfe)D(u-\halfe)-B(u+\halfe)C(u-\halfe)\\
&=&D(u+\halfe)A(u-\halfe)-C(u+\halfe)B(u-\halfe)
\end{array}
\label{eq:qdet}
\eeq
Note that the quantum determinant respects the comultiplication
$$\qdet T_1(u)T_2(u)=\qdet T_1(u)\qdet T_2(u)   $$

The simplest, one-dimensional representation of \TR\
is provided by a constant number matrix $K$
satisfying the identity
$$ [R(u),K\otimes K]=0\qquad \forall u.  $$
Actually, this condition is fulfilled for any matrix $K$
due to the $SL(2)$ symmetry of the R--matrix.
Note that the quantum determinant of $K$ coincides with its ordinary
determinant
$$\qdet K= \det K $$
We shall see in the next lecture that such representations
describe the boundary conditions for the integrable chains.

The next in complexity goes the L--operator $L(u)$ which is linear
in the spectral parameter $u$. It is constructed in terms of three
operators $S_1$, $S_2$, $S_3$ or $S_3$, $S_\pm\equiv S_1\pm iS_2$
belonging to the irreducible finite--dimensional (${\rm dim}=2l+1$)
representation of the Lie algebra $sl(2)$

\beq
  L(u)=u+\eta\sum_{\alpha=1}^3 S_\alpha\sigma_\alpha=
  \left(\begin{array}{cc}
   u+\eta S_3&\eta S_-\\
   \eta S_+&u-\eta S_3
  \end{array}\right)\qquad
   S_\pm=S_1\pm \i S_2
\label{eq:defL}
\eeq
$$[S_\alpha,S_\beta]=
     \i\sum_{\gamma=1}^3\varepsilon_{\alpha\beta\gamma}S_\gamma\qquad
\begin{array}{rcl}
  {[S_3,S_\pm]}&=&\pm S_\pm\\
  {[S_+,S_-]}&=&2S_3
\end{array}$$
$$S_1^2+S_2^2+S_3^2=S_3^2+\half(S_+S_-+S_-S_+)=l(l+1)$$
$$\qdet L(u)=(u-l\eta-\halfe)(u+l\eta+\halfe)$$

Since the R--matrix $R(u-v)$ in the quadratic relation
(\ref{eq:RTT}) determining
the algebra \TR\  depends only on the difference of the spectral
parameters, it follows immediately that shift of the spectral
parameter by a constant $L(u)\longrightarrow L(u-\d)$ is an
automorphism of \TR\  and provides thus a little bit richer
family of representations.

Now, the comultiplication property allows immediately to construct
plenty of representations of \TR\

\beq
  T(u)=KL_N(u-\d_N)\ldots L_2(u-\d_2)L_1(u-\d_1)=
  \left(\begin{array}{cc}
  A & B\\
  C & D \end{array}\right)(u)
\label{eq:KLL}
\eeq
The corresponding quantum determinant reads
$$\D(u)=\det K\prod_{n=1}^N \qdet L_n(u-\d_n)
     =\det K\prod_{n=1}^N (u-\d_n-l_n\eta-\halfe)
                    (u-\d_n+l_n\eta+\halfe)$$
To sum up, we have constructed a family of finite--dimensional
(${\rm dim}=\prod_{n=1}^N(2l_n+1)$)
representations of \TR\
$$ T(u|K\in{\rm Mat(2,2)},N\in {\bf Z}_+,
     \{l_n\in{\bf Z}/2\}_{n=1}^N,\{\d_n\in{\bf C}\}_{n=1}^N) $$
parametrized by the matrix $K$, number $N$ of L--operators,
spins $l_n$ and shifts $\d_n$.
By the reasons which will be explained in the next lecture
the corresponding quantum integrable system is called
{\it inhomogeneous XXX spin chain}.

As a matter of fact, the representations constructed
turn out to be irreducible for almost all values of $\d_n$ .
Moreover, the whole family contains all the irreducible finite--dimensional
representations of the Yangian ${\cal Y}[sl(2)]$, see
prof. M. Jimbo's lectures on the present school. Let me remark
that Yangian case corresponds to $K=1$ only.

In the next lecture we shall study these representations in more details.

\vskip 1cm
\setcounter{chapter}{2}
\setcounter{equation}{0}
\setcounter{th}{0}
\begin{center}
    \large\bf Lecture 2
\end{center}

In the last lecture we have constructed a huge family (\ref{eq:KLL})
of representations of the algebra \TR\ associated to the $sl(2)$-invariant
R-matrix (\ref{eq:RXXX}). What I am going to do now is to show that
some of the representations constructed, namely, the representations
differing by the order of factors $L_n(u-\d_n)$ in the product (\ref{eq:KLL}),
are equivalent. Obviously, it is enough to prove the statement for the
products $L_1(u)L_2(u)$ and $L_2(u)L_1(u)$. In fact, it follows from a general
theorem of Drinfeld  about the universal R-matrix claiming the equivalence
$T_1(u)T_2(u)\simeq T_2(u)T_1(u)$
for any two representations $T_{1,2}(u)$ of Yangian,
but it seems me instructive to show the equivalence manifestly on the
simplest example.

However, let us consider first more simple representations $K$ which are
absent in Drinfeld's theory.
It is obvious that the products $ K_1K_2$ and $K_2K_1 $ are not
equivalent unless $K$'s commute. However, for the products of $K$ and $L$
situation is better.
\begin{th}
 If $\det K\neq 0$ then  $KL(u)\simeq L(u)K$.
\label{th:KL}
\end{th}

{\bf Proof. } It is necessary to find such an invertible operator $\K$ in the
representation space $V$ of $L(u)$ that
$$ KL(u)=\K^{-1}L(u)K\K  $$
Let us use the $sl(2)$-invariance of the L-operator (\ref{eq:defL}):
\beq
   [L(u),S^\a+\half\s_\a]=0.
\label{eq:Linv}
\eeq
Since the matrix $K$ is invertible it can be represented as an exponent
$$ K=k_0\exp\sum_{\a=1}^3 k_\a\half\s_\a. $$
Define now $\K$ as
$$ \K\equiv k_0\exp\sum_{\a=1}^3 k_\a S^\a. $$
{}From (\ref{eq:Linv}) there  follows immediately the identity
$$ [L(u),K\K ]=0 $$
proving the theorem.

It follows from the  theorem that one can transform equivalently
 any product of $L$'s and invertible $K$'s to the form (\ref{eq:KLL})
$K$ being the product of $K$'s in the same order.

Let us return now to the products of $L$'s.
\begin{th} Let two L-operators (\ref{eq:defL}) represent the algebra
\TR\ in the spaces $V_1$ and $V_2$ and be characterised by the spins $l_{1,2}$
and shifts $\d_{1,2}$ respectively. Then an invertible operator $\R$ in the
space $V_1\otimes V_2$ intertwining the products $L_1(u-\d_1)L_2(u-\d_2)$ and
$L_2(u-\d_2)L_1(u-\d_1)$
\beq
     \R L_1(u-\d_1)L_2(u-\d_2)=L_2(u-\d_2)L_1(u-\d_1)\R
\label{eq:RLL}
\eeq
and providing therefore the equivalence of representations
    $$ L_1(u-\d_1)L_2(u-\d_2)\simeq L_2(u-\d_2)L_1(u-\d_1) $$
exists if and only if the difference $\d_{21}\equiv\d_2-\d_1$ does not
take one of the following values
\beq
   \d_{21}\neq \pm(\left|l_2-l_1\right|+1)\eta,\pm(\left|l_2-l_1\right|+2)\eta,
     \ldots, \pm(l_2+l_1)\eta
\label{eq:condLL}
\eeq
\label{th:LL}
\end{th}

{\bf Proof. }
Substituting the expression (\ref{eq:defL}) for $L$ into equation
(\ref{eq:RLL}) for $\R$ we obtain the equation
$$ \R(u-\d_1+S_1^\a\s_\a)(u-\d_2+S_2^\beta\s_\beta)=
     (u-\d_2+S_2^\beta\s_\beta)(u-\d_1+S_1^\a\s_\a)\R  $$
(here and further the summation over repeated indices is always supposed).
Separating the terms containing $u$ one notices that $\R$ should be
$SL(2)$-invariant
$$ [\R, S^\a_1+S^\a_2]=0 \qquad \forall\a$$
and depend in fact on the difference $\d_{21}=\d_2-\d_1$ only.
The remaining terms result in the equations
\beq
   \R[\d_{21}(S^\a_1-S^\a_2)+2i\varepsilon^{\a\beta\gamma}
    S^\beta_1S^\gamma_2]=
        [\d_{21}(S^\a_1-S^\a_2)-2i\varepsilon^{\a\beta\gamma}
    S^\beta_1S^\gamma_2]\R\quad \forall\a
\label{eq:equR}
\eeq

The $SL(2)$-invariance of $\R$ allows to look for $\R$ in the form
\beq
   \R=\sum_{j=\left|l_1-l_2\right|}^{l_1+l_2}
        \rho_j(\d_{21})P_j
\label{eq:expR}
\eeq
where $P_j$ are the projectors corresponding to the expansion of the tensor
product of two finite-dimensional irreducible representations of $SL(2)$
labelled by spins $l_{1,2}$ into the sum of irreducible representations
labelled by spin $j$.

Using the last expansion (\ref{eq:expR}) together with the equations
(\ref{eq:equR}) for $\R$ one obtains, after some calculation, the recurrence
relation for the eigenvalues $\rho_j(\d_{21})$ of $\R$
\beq
   \rho_{j+1}(\d_{21})=
   \frac{\d_{21}+\eta(j+1)}{\d_{21}-\eta(j+1)}\rho_j(\d_{21})
\label{eq:eigvR}
\eeq
which determines $\R$ up to an insignificant scalar factor. It remains to
notice that the necessary and sufficient condition for $\rho_j$ to be nonzero
and therefore for $\R$ to be invertible is the condition
(\ref{eq:condLL}) for $\d_{1,2}$ and $l_{1,2}$. The same condition, in fact,
ensures the irreducibility of the product $L_1(u-\d_1)L_2(u-\d_2)$
(for more details see Prof. Jimbo's lectures at the present school).
In what follows the nondegeneracy condition (\ref{eq:condLL}) is always
supposed to be fulfilled.

Let us turn now to the quantum integrals of motion $t(u)=\tr T(u)$.
The last theorem shows that the spectrum $t(u)$ does not depend on order of
$L$'s in the product (\ref{eq:KLL}). However, the explicit expression
of $t(u)$ in terms of the spin operators $S_n$ is not of course invariant
under arbitrary permutation of $S_n$.

For physical applications it is convenient to think of $S_n$ as of spins of
some atoms arranged as a one-dimensional closed chain (ring), so that it is
natural to use the periodicity condition $n+N\equiv n$ in all the formulas.
The integrable
models corresponding to the representation (\ref{eq:KLL}) of the algebra
\TR\ would look  more realistic if one could extract from $t(u)$ the
{\it local integrals of motion} that is the quantities $H^{(k)}\quad
k=1, 2, 3,\ldots$ expressible as the sums
$$ H^{(k)}=\sum_{n=1}^N H_{n,n-1,\ldots,n-k+1}^{(k)}  $$
(the periodicity $N+1\equiv1$ is supposed). The local densities
$H_{n,n-1,\ldots,n-k+1}^{(k)}$
should involve only $k$ adjacent spins $S_n, S_{n-1},\ldots, S_{n-k+1}$
$$ [H_{n,n-1,\ldots,n-k+1}^{(k)},S_m^\a]=0\qquad m>n{\ \rm or\ }m<n+k-1  $$

An important case when such local integrals exist is that of the
{\it homogeneous spin chain} corresponding to equal spins and shifts:
$l_n\equiv l,\quad \delta_n\equiv\delta=0$
$$ T(u)=KL_N(u)L_{N-1}(u)\ldots L_2(u)L_1(u) $$

The homogeneous chain has the important property of {\it translational
invariance}: the similarity transformation $U(\cdot)U^{-1}$
defined by the relations
\beq
US_n^\a U^{-1}=S_{n+1}^\a \qquad US_NU^{-1}=\K_1 S_1\K_1^{-1}
\label{eq:defU}
\eeq
leaves $t(u)$ invariant: $Ut(u)U^{-1}=t(u)$. The invariance of $t(u)$ follows
directly from the cyclic invariance of the trace and from the
theorem~\ref{th:KL}. The transformation $U(\cdot)U^{-1}$ generalizes
the ordinary translation for the periodic chain ($K=1$) to the twisted
periodic boundary condition specified by the matrix $K$. Note that,
generally speaking, $U^N\neq 1$ in contrast with the case $K=1$.

The translational invariance of $t(u)$ suggests that the local integrals
$H^{(k)}$ should also be translationally invariant:
$$ UH_{n,n-1,\ldots,n-k+1}^{(k)}U^{-1}=H_{n+1,n,\ldots,n-k+2}^{(k)}$$

The simplest one-point density is
$$ H^{(1)}_n=\ln\K_n$$
I leave the proof of  the translational invariance  $[U,H^{(1)}]=0$
and commutativity $[H^{(1)},t(u)]=0$ for you as an exercise and pass to
the problem of two-point Hamiltonian $H^{(2)}$.

The following criterion proposed by Sutherland (1970) will help us in
our search  for $H^{(2)}$.
\begin{th}
If for some translationally invariant two-point
density $H^{(2)}_{n,n-1}$ there exists a one-point translationally invariant
density $Q_n(u)$ such that
\beq
    [H_{21},L_2(u)L_1(u)]=L_2(u)Q_1(u)-Q_2(u)L_1(u)
\label{eq:Sutherland}
\eeq
then $H^{(2)}$ commutes with $t(u)$.
\label{th:Sutherland}
\end{th}
{\bf Proof. } The following calculation shows how the terms cancel
in the commutator $[H^{(2)},t(u)]$
\begin{eqnarray*}
   \lefteqn{[\ldots+H^{(2)}_{n+1,n}+H^{(2)}_{n,n-1}+\ldots,
   \ldots L_{n+1}(u)L_n(u)L_{n-1}(u)\ldots]=}  \\
 &\qquad\qquad\qquad &
   \ldots +L_{n+1}(u)L_n(u)Q_{n-1}(u)-L_{n+1}(u)Q_n(u)L_{n-1}(u)+\ldots \\
 &\qquad\qquad\qquad &
   \ldots +L_{n+1}(u)Q_n(u)L_{n-1}(u)-Q_{n+1}(u)L_n(u)L_{n-1}(u)+\ldots
\end{eqnarray*}

To finish the argument it remains to use the cyclic invariance of trace
and the translational invariance of $H^{(2)}$ to show the cancellation
of the boundary terms $n=1$ and $n=N$ which I leave to you as an exercise.

The density $H^{(2)}$ can be described in terms of the operator $\R(\d_{21})$
constructed in the theorem~\ref{th:LL}. Revising the proof of the theorem
one observes that the same $\R(\d_{21})$ intertwines operators $L_1(u)$ and
$L_2(v)$
\beq
    \R(u-v) L_1(u)L_2(v)=L_2(v)L_1(u)\R(u-v)
\label{eq:RLuLv}
\eeq
Using the formula (\ref{eq:eigvR}) it is easy to see that
$\rho_{j+1}(0)=-\rho_j(0)$ and hence $\R(0)$ is
proportional to the permutation operator $\PP$ in $V\otimes V$.
Let us normalise $\R(u)$ multiplying it by appropriate scalar factor
such that $\R(0)=\PP_{12}$. Then the  wanted
expression for $H^{(2)}_{21}$ is given by
$H^{(2)}_{21}=\dot{\R}(0)\PP_{12}$
where $\dot{\R}$ stands for the derivative $d\R/du$.

To prove the statement let us differentiate (\ref{eq:RLuLv})
with respect to $u$ at $u=v$. Multiplying the result
$$ H_{21}\PP_{12}L_1(u)L_2(u)+\PP_{12}\dot{L}_1(u)L_2(u)
  =L_2(u)\dot{L}_1(u)\PP_{12}+L_2(u)L_1(u)H_{21}\PP_{12} $$
from the right hand side by $\PP_{12}$ and using the obvious identities
$\PP_{12}\PP_{12}=1$,\  $\PP_{12}L_1\PP_{12}=L_2$,\
$\PP_{12}L_2\PP_{12}=L_1$ we arrive at the Sutherland equation
(\ref{eq:Sutherland}) for $ Q_n(u)=\dot{L}_n(u) $.

Using the formula (\ref{eq:eigvR}) for the eigenvalues of $\R$ one
can derive the eigenvalue expansion for $H^{(2)}_{21}$
$$ H^{(2)}_{21}={\rm const}+{\rm coeff}\times
            \sum_{j=\left|l_1-l_2\right|}^{l_1+l_2}P_j
           \sum_{k=1}^j\frac{1}{k}  $$

One can show that the higher local integrals $H^{(3)}$ \etc\ also
exist for the homogeneous XXX spin chain.
The most popular method to produce them belongs to Baxter (1972)
and L\"{u}scher (1976) and gives an expression for $H^{(k)}$ in terms
of higher derivatives of $\R(u)$ at $u=0$.
I'll not go into details of the calculation which are
well described in the literature and simply reproduce the result.

Let us construct first the quantities $t^{(k)}(u)$ which are easier
to describe by words then to write a compact formula. Let $\tt(u)=
t(u+\halfe)$.
To construct
$t^{(k)}(u)$ take the product $\tt(u)\tt(u+\eta)\ldots \tt(u+(k-1)\eta)$.
Then subtract all possible terms obtained from it after replacing
two adjacent factors $t(u+k\eta)t(u+(k+1)\eta)$ by
$\DD(u+(k+\half)\eta)$ where
$\DD(u)=\D(u+\halfe)$ is obtained from
the quantum determinant (\ref{eq:qdet}) by the same shift as $\tt(u)$
from $t(u)$.  Then add all terms obtained from the original product
after replacing in the same way two pairs of $\tt$'s by $\DD$'s and
continue changing sign at each step until the $t$'s will be exhausted.
The result is $t^{(k)}(u+(k-1)\halfe)$. For example
\begin{eqnarray*}
 t^{(1)}(u)&=&\tt(u) \\
 t^{(2)}(u+\halfe)&=&\tt(u)\tt(u+\eta)-\DD(u+\halfe) \\
 t^{(3)}(u+\eta)&=&\tt(u)\tt(u+\eta)\tt(u+2\eta)
      -\DD(u+\halfe)\tt(u+2\eta)-\tt(u)\DD(u+\frac{3\eta}{2}) \\
 t^{(4)}(u)&=&\tt(u)\tt(u+\eta)\tt(u+2\eta)\tt(u+3\eta)
        -\DD(u+\halfe)\tt(u+2\eta)\tt(u+3\eta)\\
      &&  -\tt(u)\DD(u+\frac{3\eta}{2})\tt(u+3\eta)
        -\tt(u)\tt(u+\eta)\DD(u+\frac{5\eta}{2}) \\
      && +\DD(u+\halfe)\DD(u+\frac{5\eta}{2})
\end{eqnarray*}
and so on.

It turns out that if $l$ is the spin of the homogeneous XXX spin chain
then $t^{(2l)}(u)$ taken at $u=0$ is proportional to the generalised
translation operator $U$ defined previously (\ref{eq:defU}). Finally,
the local integrals of motion $H^{(k)}$ are obtained as the coefficients of the
power series
\beq
    \tau(u)\equiv
   \ln U^{-1}t^{(2l)}(u)=\sum_{k=1}^\infty \frac{u^k}{k!} H^{(k+1)}
\label{eq:deftau}
\eeq

In conclusion, I want to show you a simple method for generating higher
integrals which is not so widely known. The method is based on the concept
of {\it master symmetry} which can be defined as an operator $\BB$
producing one integral of motion from another
\beq
    [\BB,H^{(k)}]=H^{(k+1)}
\label{eq:comBH}
\eeq

Tetelman (1982) has shown that for the integrable spin chains the
master symmetry $\BB$ can be taken as the following sum (``boost'' operator)
$$\BB=\sum_n nH^{(2)}_{n,n-1} $$

Strictly speaking, the above formula makes sense only for the infinite chain
since the linearly growing coefficient $n$
contradicts obviously the periodicity of the
spin chain. Nevertheless, in the periodic case
 one can use it quite formally in commutators
which ``differentiate'' the linear dependence on $n$ and yield periodic
expressions.

The possibility to use $\BB$ as a master symmetry is based on
the obvious identity
\beq [\BB,U^n]=nU^nH^{(2)}
\label{eq:comBU}
\eeq
and the identity
\beq
   [\BB,t(u)]=\dot{t}(u)
\label{eq:comBt}
\eeq
which is proven in the same way as the theorem~\ref{th:Sutherland}
with the only difference that due to the factor $n$ in the definition of $\BB$
the terms do not cancel completely but add up to $\dot{t}(u)$
\begin{eqnarray*}
   \lefteqn{[\ldots+(n+1)H^{(2)}_{n+1,n}+nH^{(2)}_{n,n-1}+\ldots,
   \ldots L_{n+1}(u)L_n(u)L_{n-1}(u)\ldots]=} \\
 &\qquad &  \ldots +nL_{n+1}(u)L_n(u)\dot{L}_{n-1}(u)-
          nL_{n+1}(u)\dot{L}_n(u)L_{n-1}(u)+\ldots \\
&\qquad &  \ldots +(n+1)L_{n+1}(u)\dot{L}_n(u)L_{n-1}(u)-
          (n+1)\dot{L}_{n+1}(u)L_n(u)L_{n-1}(u)+\ldots
\end{eqnarray*}

Using the identities (\ref{eq:comBU}) and (\ref{eq:comBt}) one obtains the
commutation relation
$$ [\BB,\tau(u)]=\dot{\tau}(u)-H^{(2)} $$
between $\BB$ and the generating function (\ref{eq:deftau}) of $H^{(k)}$.
Expanding the last equality in powers of $u$ one arrives at the wanted
relation (\ref{eq:comBH}).

\vskip 1cm
\setcounter{chapter}{3}
\setcounter{equation}{0}
\setcounter{th}{0}
{\samepage
\begin{center}
    \large\bf Lecture 3
\end{center}

\nopagebreak
Now we are in a position to attack the main problem of QISM ---
the spectral analysis of the commuting family $t(u)$, see Step 3 of
the general scheme given in the Lecture 1. Traditionally, various
methods of solving the problem have the common name: Bethe Ansatz (BA),
with the corresponding adjective: coordinate BA, algebraic BA,
analytic BA, functional BA, nested BA \etc. In the last two lectures I'll
touch briefly the algebraic BA (ABA) described at length in Prof.
L. D. Faddeev's Nankai lectures (1987)
and concentrate on the new variant of BA --- the functional BA.\par}
The ABA works for the highest vector  representations $T(u)$ of the Yangian
${\cal Y}[sl(2)]$
that is the representations possessing the heighest (vacuum) vector
$\vac$ which is
annihilated by the off-diagonal element $C(u)$ of the matrix $T(u)$
and which is an eigenvector of its diagonal elements $A(u)$ and $D(u)$
$$C(u)\vac=0\qquad
  \begin{array}{rcl}
   A(u)\vac&=&\D_-(u)\vac\\
   D(u)\vac&=&\D_+(u)\vac\end{array}\qquad\forall u$$

It is easy to show that the eigenvalues $\D_\pm(u\mp\halfe)$
factorize the quantum determinant $\D(u)$ of $T(u)$
$$\D_+(u-\halfe)\D_-(u+\halfe)=\D(u)$$

Let us look now for the eigenvectors of $t(u)$ in the form
$$\left|v_1,v_2,\ldots,v_M\right>=\prod_{m=1}^M B(v_m)\vac$$
Then it is possible to show that the eigenvalue problem
$$t(u)\left|v_1,v_2,\ldots,v_M\right>=\tau(u)\left|v_1,v_2,\ldots,v_M\right>$$
is equivalent to the set of Bethe equations
\beq
 \frac{\D_+(v_m)}{\D_-(v_m)}=
  \prod_{\scriptstyle n=1\atop\scriptstyle n\neq m}^M
  \frac{v_m-v_n-\eta}{v_m-v_n+\eta}\qquad m=1,\ldots,M
\label{eq:Bethe}
\eeq
the corresponding eigenvalue $\tau(u)$ of $t(u)$ being
\beq
\tau(u)=\D_-(u)\prod\limits_{m=1}^M \frac{u-v_m-\eta}{u-v_m}+
        \D_+(u)\prod\limits_{m=1}^M \frac{u-v_m+\eta}{u-v_m}
\label{eq:eigvtau}
\eeq
Note that one can obtain the Bethe equations (\ref{eq:Bethe}) from the
formula (\ref{eq:eigvtau}) for $\tau(u)$ by taking residue at $u=v_m$ and
using the smoothness of the polynomial $\tau(u)$.
The last equation (\ref{eq:eigvtau}) is in turn equivalent to the linear
finite-difference spectral problem (Baxter's equation)

\beq
  \tau(u)Q(u)=\D_+(u)Q(u+\eta)
                   +\D_-(u)Q(u-\eta)
\label{eq:Baxter}
\eeq
for the polynomial $Q(u)$ determined by its zeroes $\{v_m\}_{m=1}^M$
$$Q(u)\equiv\prod_{m=1}^M(u-v_m)$$
We shall return to the Baxter's equation when speaking about the
Functional Bethe Ansatz.

Let us apply the ABA construction to our favorite example ---
the inhomogeneous XXX magnetic chain. It is a simple excersize to show that the
vacuum vector exists if and only if the matrix $K$ determining the
boundary condition is triangular $(K_{21}=0)$. For the sake of simplicity
we shall assume $K$ to be diagonal
$$ K=\left(\begin{array}{cc}
        \xi_- & 0\\
        0   & \xi_+
     \end{array}\right)        $$
In that case
the vacuum $\vac$ is the common heighest vector of the local
representations $S_n$ of $su(2)$
           $$S_n^-\vac=0\qquad S_n^3\vac=-l_n\vac\qquad
           \forall n\in\{1,\ldots,N\}$$
The corresponding eigenvalues of $A(u)$ and $D(u)$ are polynomials
of degree $N$
$$\D_\pm(u)=\xi_\pm\prod_{n=1}^N (u-\d_n\mp l_n\eta)$$

The ABA proved to be quite powerful method which helped to solve such
models as sine-Gordon, XXX and XXZ magnets and others.
However, being restricted to the heighest vector representations, it
fails for representations which do not satisfy this condition.
In particular, such interesting integrable models as sinh-Gordon,
Toda chain, quantum tops fall beyond the reach of ABA. One should mention
also the so-called {\it completeness problem} that is the question if
the Bethe eigenstates $\left|v_1,v_2,\ldots,v_M\right>$ are complete.
There are simple examples when this is not the fact and some special
investigation is needed in order to gain the missing eigenstates.

In the rest of this lecture and in the last one I'll describe an
alternative method, the Functional Bethe Ansatz (FBA), which is free
of the restrictions inherent in ABA.

The Functional Bethe Ansatz was born from a fortunate marriage
of two ideas. The first one is the central idea of QISM that is
using quadratic R matrix algebras (Yangians \etc) to describe
the dynamic symmetry of integrable systems. The second one is
the very old idea of \sov. So, before I'll
proceed to FBA itself, let me remind you some elementary
facts concerning the \sov.

Briefly, the \sov\ is a reduction of a multidimensional
spectral problem to a system of one-dimensional (multiparameter)
spectral problems. The simplest example of \sov\ is provided by the
Cartesian coordinates on the plane ${\bf R}^2\ni(x_1,x_2)$.

Let us introduce two commuting differential operators of second
order
$$D_1=\frac{\partial^2}{\partial x_1^2}\qquad
  D_2=\frac{\partial^2}{\partial x_2^2}$$
$$[D_1,D_2]=0$$
which can be thought of as the integrals of motion of some integrable
quantum system. It follows from the commutativity that the operators $D_i$
possess common eigenfunctions
$$\left\{\begin{array}{c}
      D_1\Phi=\lambda_1\Phi\\
      D_2\Phi=\lambda_2\Phi\end{array}\right.\qquad  $$

It is well known that the eigenfunctions factorize
$$  \Phi(x_1,x_2)=X_1(x_1)X_2(x_2)$$
the original two-dimensional spectral problem being thus reduced
to two independent one-dimensional spectral problems
$$\left\{\begin{array}{c}
      X_1^{\prime\prime}=\lambda_1X_1\\
      X_2^{\prime\prime}=\lambda_2X_2\end{array}\right.$$

Less trivial example is presented by the parabolic coordinates
$(y_1,y_2)$
$$\left\{\begin{array}{rcl}
   x_1&=&\frac{\displaystyle y_1^1-y_2^2}{\displaystyle 2}\\
   x_2&=&y_1y_2\end{array}\right.$$

The corresponding commuting second-order differential operators are
$$\begin{array}{rclcl}
  \D&=&\frac{\partial^2}{\partial x_1^2}+
           \frac{\partial^2}{\partial x_2^2}&=&
  \frac{1}{y_1^2+y_2^2}\left(\frac{\partial^2}{\partial y_1^2}+
           \frac{\partial^2}{\partial y_2^2}\right)\\
   D&=&-2x_1\frac{\partial^2}{\partial x_2^2}+
        2x_2\frac{\partial^2}{\partial x_1\partial x_2}+
            \frac{\partial}{\partial x_1}&=&
      \frac{1}{y_1^2+y_2^2}\left(y_2^2\frac{\partial^2}{\partial y_1^2}-
           y_1^2\frac{\partial^2}{\partial y_2^2}\right)\end{array}$$
As previously, their common eigenfunctions
$$   \D\Phi=\lambda\Phi \qquad
   D\Phi=\mu\Phi$$
factorize
$$  \Phi(y_1,y_2)=Y_1(y_1)Y_2(y_2)$$

The corresponding separated equations are
$$\left\{\begin{array}{c}
  Y_1^{\prime\prime}-(\lambda y^2+\mu)Y_1=0\\
  Y_2^{\prime\prime}-(\lambda y^2-\mu)Y_2=0\end{array}\right.$$

Note that, in contrast with the previous example,  the spectral parameters
$\lambda$ and $\mu$ cannot be
decoupled, and both equations should be solved together.
This situation of the so called
{\it Multiparameter spectral problem} is typical for the \sov\
in the general case. Note that the Baxter's equation (\ref{eq:Baxter})
arisen in ABA is another example of (finite-difference) multiparameter
spectral problem.

In the above examples the \sov\ was produced by
a pure change of coordinates. In the most general case, however,
the separation  should be obtained via generic canonical
transformation involving not only coordinates but also momenta.
The simplest example of this kind is the Goryachev-Chaplygin top.

The quantum GC top is described in terms of the operators
$J_\alpha, x_\alpha\quad \alpha=1,2,3$ belonging to  the $e(3)$ Lie algebra
$$\begin{array}{rcl}
   {[J_\alpha,J_\beta]}&=&-\i\varepsilon_{\alpha\beta\gamma} J_\gamma\\
   {[J_\alpha,x_\beta]}&=&-\i\varepsilon_{\alpha\beta\gamma} x_\gamma\\
   {[x_\alpha,x_\beta]}&=&0\end{array}$$
The values of the Casimir operators are assumed to be fixed
$$\sum_{\alpha=1}^3x_\alpha^2=1\qquad
   \sum_{\alpha=1}^3x_\alpha J_\alpha=0$$
Note that the second constraint is crucial for integrability.

The Hamiltonian of the model is
\beq
   H=\half(J_1^2+J_2^2+4J_3^2)-bx_1=
     \half(J^2+3J_3^2)-bx_1
\label{eq:Htop}
\eeq
where $b$ is a parameter (magnitude of the external field).

The second integral of motion commuting with $H$ is
\beq
  G=2J_3(J^2-J_3^2+\frac{1}{4})+b(x_3J_1+J_1x_3)
\label{eq:Gtop}
\eeq
(note the quantum correction $\frac{1}{4}$).

The separated variables $(u_1,u_2)$ in the classical case were
found by Chaplygin himself
$$\begin{array}{c}
  u_1=J_3+\sqrt{J^2}\\
  u_2=J_3-\sqrt{J^2}\end{array}$$

Their quantum generalization has been found few years ago by
Komarov. They are defined as the ``operator roots'' of the quadratic
polynomial with commuting coefficients
$$u^2-2J_3u-(J^2-J_3^2+\frac{1}{4})$$
(note the quantum correction $\frac{1}{4}$). The common spectrum of
$(u_1,u_2)$ turns  out to form an equidistant square lattice
$${\rm spec}(u_1,u_2)=\{1\mp\half+2n_1, -1\pm\half-2n_2\}$$

The common eigenfunctions of $H$ and $G$ factorize in the $u$-representation
$$\Phi(u_1,u_2)=\phi(u_1)\phi(u_2)$$

The corresponding one-dimensional separated equation is the same for
$u_1$ and $u_2$
$$\tau(u)\phi(u)=d(u+1)\phi(u+2)+d(u-1)\phi(u-2)$$
where
$$\tau(u)=u^3-2(H+\frac{1}{8})u-G$$
$$d(u)=b\sqrt{u^2-\frac{1}{4}}$$

The example of GC top shows how nontrivial  can be the task of finding
the \sov. There is no general recipe guaranteeing success for any
given integrable model. Fortunately, for the integrable systems
subject to QISM such a recipe, to which I have given the name
``Functional Bethe ansatz (FBA)'', exists, at least in the case of
the Yangian ${\cal Y}[sl(2)]$. The success of FBA lies in using
the algebraic machinery
of QISM which provides a powerful tool for finding the separated variables.

So, I am starting the detailed exposition of FBA.
Let me remind you that throughout these lectures we consider the algebra
\TR\ associated to the R matrix of the XXX model. Suppose that
the $2\times2$ matrix $T(u)$ is a polynomial in the spectral parameter $u$
$$ T(u)=
             \left(\begin{array}{cc}
                A(u) & B(u)\\
                C(u) & D(u)\end{array}\right)    =
\sum_{n=0}^N u^nT_n=
        \sum_{n=0}^N u^n
             \left(\begin{array}{cc}
                A_n & B_n\\
                C_n & D_n\end{array}\right)
         $$
and defines a representation of \TR\ in a finite-dimensional space $W$.

The recipe which constitutes the heart of FBA can be expressed in few words.
Take the polynomial $B(u)$ having commuting operator coefficients
(the commutativity
$$ [B(u),B(v)]=0\qquad\forall u,v  $$
is a simple consequence of the basic quadratic relation
(\ref{eq:RTT}) for the T matrix).
Then the operator zeroes $x_n$ of $B(u)$
provide the wanted separated variables. The rest of my lectures will
be devoted to the deciphering and adjustment of this obscure remark.

The first problem which arises immediately is to give the precise meaning
to the roots of the operator polynomial $B(u)$.
In order to simplify the task and to avoid troubles with degenerate
cases I'll impose few conditions on the representation $T(u)$.
The first condition is always fulfilled for irreducible representations
of \TR.
\begin{cond}
  The senior coefficient  $T_N$ is a number matrix. The quantum determinant
$\D(u)$ of $T(u)$ is a number-valued function.
\end{cond}

The second one is a nondegeneracy condition. It ensures that $B(u)$
and $\D(u)$ are polynomials of the maximal degree.
\begin{cond}
 $$ B_N\neq0\qquad \det T_N\neq 0   $$
\end{cond}

Though it is not yet clear what are the operator roots of $B(u)$ their
symmetric polynomials $\bb_n$ are easy to define
$$ \bb_n\equiv(-1)^n B_{N-n}/B_N\qquad n=1,2,\ldots,N $$

Using $\bb_n$ one can express $B(u)$ as
$$ B(u)=B_N(u^N-\bb_1u^{N-1}+\bb_2u^{N-2}-\cdots)$$

It follows from the commutativity of $B(u)$ that the operators
$\{\bb_n\}_{n=1}^N$ also commute. Let $\B$ be their common spectrum.
$$ \B={\rm spec} \{\bb_n\}_{n=1}^N$$

The following condition of semisimplicity is the most restrictive one.
Fortunately, it holds for many interesting models.

\begin{cond}
  The operators $\{\bb_n\}_{n=1}^N$ have a complete set of common
eigenfunctions and to every point $\b=(b_1,\ldots,b_N)\in\B\subset\C^N$
there corresponds only one eigenfunction.
\end{cond}

It follows from the last condition that the representation space $W$
is isomorphic to the space $\Fun\B  $ of functions on the finite
set $\B\subset\C^N$. Of course, the isomorphism is not unique. It
is defined up to multiplication by some nonzero function on $\B$.
So, let us suppose that some isomorphism is fixed
and the operators $\bb_n$ are realized as the operators of multiplication
by the corresponding coordinates $b_n$ in $\C^N$
$$ \b=(b_1,\ldots,b_N)\in\C^N\qquad f\in\Fun\B$$
$$ [\bb_nf](\b)=b_nf(\b)$$
In what follows in such cases I
will not make difference between the operators  like $\bb_n$ and
the corresponding functions  like $b_n$ and always will omit hats over
operators (see later operators $x_n$).

Since $b_n$ are the symmetric polynomials of the roots of $B(u)$
which are yet to be defined we are led to consider the mapping
$$ \Theta:\C^N\rightarrow\C^N:\x\rightarrow\b  $$
given by the formula
$$ b_n(\x)=s_n(\x)  $$
where $s_n(\x)$ is the elementary symmetric polynomial of
degree $n\in\{1,\ldots,N\}$ of $N$ variables $\{x_n\}_{n=1}^N$.
$$ \begin{array}{rcl}
      s_1(\x)&=&x_1+x_2+\cdots+x_N\\
      s_2(\x)&=&x_1x_2+\cdots\\
      &\cdots&\\
      s_N(\x)&=&x_1x_2\ldots x_N
   \end{array}                                    $$
Note that the pre-image $\Theta^{-1}(\B)$ is symmetric under
permutations of the coordinates $\{x_n\}_{n=1}^N$.

And finally, the very last condition.
\begin{cond}
  Pre-image $\X=\Theta^{-1}(\B)$ contains no multiple points that is
each $\b\in\B$ has exactly $N!$ pre-images.
\end{cond}
In principle this condition can be omitted but in that case one is urged
to use rather sophisticated language of jets in order to work with the
multiple points.

It follows from the last condition that the mapping $\Theta$ induces
isomorphism between $W=\Fun \B$ and the space $\Sym\,\Fun\X$.
Now it is easy to define the operator roots $x_n$ of $B(u)$
as multiplication operators in the {\it extended representation space}
$$ \tilde{W}=\Fun \X\supset W=\Sym\,\Fun\X $$
It is obvious that ${\rm spec}\{x_n\}_{n=1}^N=\X$ and that
\beq
   B(u)=B_N\prod_{n=1}^N(u-x_n)
\label{eq:Bexp}
\eeq

For the last equality to be correct its right hand side should be restricted
to the space $W$ since $B(u)$ is originally is defined only on $W$.
However, the same equality can be considered as definition of the natural
extension of $B(u)$ from $W$ to $\tilde{W}$.

Working with $\tilde{W}$ one should always keep in mind that this
is a ``nonphysical space'' and all the final results should use
only the original space $W=\Fun \B$=$\Sym\,\Fun\X$.

The operators $x_n$ being defined the next problem is to calculate
the expression for the commuting integrals of motion
$t(u)=\tr T(u)=A(u)+D(u)$ in the $x$-representation and to observe the
resulting separation of variables. This task will be solved in several
steps. First, let us introduce the ``momenta'' conjugated to the
``coordinates'' $x_n$. Consider the operators $X_n^\pm$
\beq
   X^-_n=\sum_{p=1}^N x_n^p A_n\equiv [A(u)]_{u=x_n}
\label{eq:defXm}
\eeq
\beq
   X^+_n=\sum_{p=1}^N x_n^p D_n\equiv [D(u)]_{u=x_n}
\label{eq:defXp}
\eeq
which are obtained from the polynomials $A(u)$ and $D(u)$ as their
values at the ``points'' $u=x_n$. The substitution of operator
values for $u$ will be defined correctly if one prescribes some rule
for operator ordering. Here the ordering of $x$'s to the left is chosen.
I shall call it ``substitution from the left''.
Note that the operators $X_n^\pm$ act from $W$ to $\tilde{W}$.

It is convenient to introduce the shift operators $E_n^\pm=e^{\pm\eta\partial/
\partial x_n}$
$$ E_n^\pm:\C^N\rightarrow\C^N:
   (x_1,\ldots,x_n,\ldots,x_N)\rightarrow(x_1,\ldots,x_n\pm\eta,\ldots,x_N)
$$

It turns out that the operators $X_n^\pm$ have nice commutation relations
with $x_n$.
\begin{th}
\beq
   X_m^\pm x_n=(x_n\pm\eta\d_{mn})X_m^\pm  \qquad\forall m,n
\label{eq:Xx}
\eeq
\label{th:XX}
\end{th}

{\bf Proof.} Consider the following commutation relation between
$A(u)$ and $B(u)$ which can be extracted from the basic commutation relation
(\ref{eq:RTT})
$$ (u-v)A(v)B(u)+\eta B(v)A(u)=(u-v+\eta)B(u)A(v)  $$

Now substitute $x_n$ for $v$ from the left. The second term in the left hand
side will vanish since $B(x_n)=0$.
$$ (u-x_n)X_n^-B(u)=(u-x_n+\eta)B(u)X_n^-  $$

After substituting the expansion (\ref{eq:Bexp}) for $B(u)$, cancelling
the factors $B_N$ and $(u-x_n)$, and expanding both sides in powers of $u$
we obtain the relation
\beq
    X_n^-s(\x)=s(E_n^-\x)X_n^-
\label{eq:Xs}
\eeq
first for any elementary symmetric and hence for any symmetric
polynomial $s(\x)$. Note that, since $\X$ is a finite set, $s(\x)$
can be in fact any symmetric function on $\X$.

Quite similarly, the commutation relation between $X_m^+$ and
$x_n$ is obtained from the identity
$$ (u-v)D(u)B(v)+\eta D(u)B(v)=(u-v+\eta)B(v)D(u)  $$

The next natural step would be establishing the commutation relations
between $X^\pm$'s. However, it cannot be done immediately because of
the fact that the operators $X_n^\pm$ act by definition from
$W$ to $\tilde{W}$. So, we need first to extend them from $W$ to
$\tilde{W}$ as we did it with $B(u)$. To this end, consider the
constant function $\omega$ on $\X$
$$ \omega(\x)=1\qquad \forall \x\in\X $$
Obviously, $\omega$ is symmetric under permutations and thus belongs
to $\Sym\,\Fun\X=W$. Consider now the functions $\D_n^\pm$ on $\X$
$$\D_n^\pm(\x)=[X_n^\pm \omega](\x) \qquad\forall \x\in\X$$
defined as the images of $\omega$ for the operators $X_n^\pm$.

Let us show now that due to the commutation relations (\ref{eq:Xx})
between $X_m^\pm$ and $x_n$
the functions $\D_n^\pm$ on $\X$ determine uniquely the action
of $X_n^\pm$ on any vector $s$ of $W$. Note that any vector $s\in W$
which is identified due to the isomorphism $W=\Fun\X$ with some
symmetric function
$s(x_1,\ldots,x_N)$ can be created from the cyclic vector $\omega$
by the operator $\hat{s}=s(\hat{x}_1,\ldots,\hat{x}_N)$. Then, applying
$X_n^\pm$ to $s$ and using the identity
(\ref{eq:Xs}) for the function $s(\x)$ we obtain
$$ [X_n^\pm s](\x)=[X_n^\pm \hat{s}\omega](\x)=s(E_n^\pm\x)[X_n^\pm\omega](\x)
   =s(E_n^\pm\x)\D_n^\pm(\x)      $$

Now we can consider the last equality as the definition of action of
$X_n^\pm$ on {\it arbitrary}, not necessarily symmetric,
function $s\in\Fun\X=\tilde{W}$. Obviously the commutation relations
(\ref{eq:Xx})
with $x$'s will still be valid for $X^\pm$'s thus extended on $\tilde W$.
Consequentely, the relation (\ref{eq:Xs})
 is now valid for any (not necessarily symmetric) function $s(\x)$.
In other words, the extended operators $X_n^\pm$ can be expressed
in terms of multiplication and shift operators
\beq
 X_n^\pm=\D_n^\pm E_n^\pm
\label{eq:exprX}
\eeq

Now we can calculate the commutation relations for extended $X^\pm$'s.

\begin{th}
\begin{eqnarray}
    [X_m^\pm,X_n^\pm]&=&0  \qquad\forall m,n \nonumber \\
 {[}X_m^+,X_n^-{]}&=&0  \qquad\forall m,n \quad m\neq n
\label{eq:XX} \\
   X_n^\pm X_n^\mp&=&\D(x_n\pm\halfe)  \qquad\forall n \nonumber
\end{eqnarray}
\end{th}

{\bf Proof.} Let us prove first the commutativity of $X_n^-$'s.
The assertion  being obvious for $m=n$, it is enough to consider
the case $m=1$, $n=2$.
Take the product $A(u)A(v)$ and substitute in it $u=x_1$ and
$v=x_2$
$$ [A(u)A(v)]_{\scriptstyle u=x_1\atop\scriptstyle v=x_2}=
   \sum_{n_1n_2}x_1^{n_1}x_2^{n_2}A_{n_1}A_{n_2}=\ldots$$

Then, using the commutativity of $x$'s and the definition of $X_1^-$
we obtain
$$    \ldots=\sum_{n_1n_2}x_2^{n_2}x_1^{n_1}A_{n_1}A_{n_2}
      =\sum_nx_2^nX_1^-A_n=\ldots    $$

Finally, using commutativity of $x_2$ and $X_1^-$ and the definition of
$X_2^-$ we arrive at $X_1^-X_2^-$
$$\ldots=\sum_n X_1^-x_2^nA_n=X_1^-X_2^- $$

In the same way, starting from $A(v)A(u)$ one obtains $X_2^-X_1^-$.
Since $A(u)$ and $A(v)$ commute due to (\ref{eq:RTT})
 so do $X_1^-$ and $X_2^-$,
the first assertion of the theorem being thus proven. Quite analogously,
the commutativity of $X^+$'s follows from the commutativity of $D(u)$.
Similarly, commutativity of $X_m^+$ and $X_n^-$ for $m\neq n$ is derived from
the identity
$$ (u-v)D(u)A(v)+\eta B(u)C(v)=(u-v)A(v)D(u)+\eta B(v)C(u) $$

Consider now one of equalities (\ref{eq:qdet}) for the quantum determinant
$$ \D(u-\halfe)=A(u)D(u-\eta)-B(u)C(u-\eta)  $$

Substituting $u=x_n$ from the left leaves only the first term in the right hand
side. Proceeding in the same way as above we obtain

$$  \begin{array}{rcl}
\D(x_n-\halfe)&=&
   [A(u)D(u-\eta)]_{u=x_n}=\sum\limits_{p,q}x_n^p(x_n-\eta)^qA_pD_q=
   \sum\limits_{p,q}(x_n-\eta)^qx_n^pA_pD_q       \\
   &=&\sum\limits_q(x_n-\eta)^qX_n^-D_q=\ldots
   \end{array}                          $$

Now it remains to use the commutation relation (\ref{eq:Xx})
 between $X_n^-$ and
$x_n$ to arrive at the wanted result
$$   \ldots=\sum_qX_n^-x_n^qD_q=X_n^-X_n^+ $$

The analogous equality for $X_n^+X_n^-$ is obtained in the same way
starting from the identity
$$ \D(u+\halfe)=D(u)A(u+\eta)-B(u)C(u+\eta)  $$

I propose as a home exercise to look  once more the proof of the
theorem through and to trace where the extended definition of $X^\pm$'s was
used and where the original one (\ref{eq:defXm}), (\ref{eq:defXp}).

Let us write down the whole set of commutation relations between
$x$'s and $X^\pm$'s

$$ \begin{array}{rclcl}
    [x_m,x_n]&=&0 & \qquad & \forall m,n \\
    X_m^\pm x_n&=&(x_n\pm\eta\d_{mn})X_m^\pm & \qquad &\forall m,n \\
    {[}X_m^\pm,X_n^\pm{]} &=&0  & \qquad & \forall m,n \\
    {[}X_m^+,X_n^-{]} &=&0  & \qquad & \forall m,n \quad m\neq n\\
     X_n^\pm X_n^\mp&=&\D(x_n\pm\halfe) & \qquad & \forall n
   \end{array}                                               $$

It is quite natural to inquire about the representation theory for
the associative algebra \XD\ defined by the generators
$\{x_n,X_n^\pm\}_{n=1}^N$ and the above relations, and labelled by
the polynomial $\D(u)$ of degree $2N$.

In the case of finite-dimensional representations such that the spectrum
$\X$ of the operators $\{x_n\}_{n=1}^N$ is simple and has no multiple
points, as throughout my lecture due to the Conditions 1--4,
the problem of constructing a representation of \XD\  is equivalent
to that of finding the functions $\{\D_n^\pm\}_{n=1}^N$ on $\X$
satisfying certain relations which follow from the relations (\ref{eq:XX})
between $X^\pm$'s
\begin{eqnarray}
    \D_m^\pm(\x)\D_n^\pm(E_m^\pm\x)
   &=&\D_n^\pm(\x)\D_m^\pm(E_n^\pm\x)
   \qquad\forall m,n\quad\forall \x\in\X \nonumber \\
    \D_m^+(\x)\D_n^-(E_m^+\x)
   &=&\D_n^-(\x)\D_m^+(E_n^-\x)
   \qquad\forall m,n\quad m\neq n\quad\forall \x\in\X
\label{eq:Dcond1} \\
 \D_n^\pm(\x)\D_n^\mp(E_n^\pm\x)&=&\D(x_n\pm\halfe)
   \qquad\forall n\quad\forall \x\in\X  \nonumber
\end{eqnarray}

Strictly speaking, the above relations are not defined when the shift
$E_n^\pm$ moves the point $\x$ out of the set $\X$. It means that the
functions $\{\D_n^\pm\}_{n=1}^N$ must vanish on the boundary $\X_n^\pm$
of the set $\X$ with respect to the shift $E_n^\pm$
\begin{eqnarray}
&  \D_n^\pm(\x)=0\quad \forall \x\in\X_n^\pm & \nonumber \\
&&  \label{eq:Dcond2} \\
&  \X_n^\pm=\{\x\in\X\mid E_n^\pm\x\in(\C\setminus\X)\} & \nonumber
\end{eqnarray}
As a consequence, the function $\D(u)$ must vanish
on the set
$$\bigcup_{n=1}^N\left((\X_n^++\halfe)\cup(\X_n^--\halfe)\right)$$

Irreducibility and finite-dimensionality must put severe restrictions
on the set $\X$ and the functions $\D(u)$, $\D_n^\pm$. It seems highly
probable that a kind of Stone-von Neumann theorem should exist for the
algebra \XD\  that is there should exist essentially unique irreducible
representation for every allowed $\D(u)$.

Let us consider a simple example of irreducible representation for \XD.
Let numbers $\{\l_n^\pm\}_{n=1}^N$ be such that
$(\l_n^+-\l_n^-)/\eta=2l_n$ is a positive integer $\forall n$.
Let the sets $\L_n$
be the equidistant (step $\eta$) strings connecting $\l_n^\pm$
\beq
 \L_n=\{\l_n^-,\l_n^-+\eta,\ldots,\l_n^+-\eta,\l_n^+\}\qquad
      \left|\L_n\right|=2l_n+1
\label{eq:defLam}
\eeq
Let us suppose, in addition, that the sets $\L_n$ do not intersect
\beq
 m\neq n \quad\Rightarrow\quad \L_m\cap\L_n=\emptyset
\label{eq:nondegL}
\eeq

Define now $\X$ as the parallelepiped
\beq
 \X=\L_1\times\L_2\times\cdots\times\L_N\qquad
   \left|\X\right|=\prod_{n=1}^N(2l_n+1)
\label{eq:setX}
\eeq
and the functions  $\D_n^\pm$, $\D(u)$ as
$$ \D_n^\pm(\x)=\D_\pm(x_n)  $$
$$  \D_\pm(u)=\xi_\pm\prod_{n=1}^N(u-\l_n^\pm) \qquad  \xi_\pm\neq 0$$
\beq
  \D(u)=\D_-(u+\halfe)\D_+(u-\halfe)
\label{eq:defD}
\eeq
where $\xi_\pm$ are arbitrary nonzero numbers.

\begin{th}
The functions $\D_n^\pm$ define an irreducible representation of
the algebra \XD\  in the space $\Fun\X$.
\label{th:irred}
\end{th}

{\bf Proof.}
One can verify easily that the functions $\D_n^\pm$ satisfy the
conditions (\ref{eq:Dcond1}), (\ref{eq:Dcond2})
 and hence define a representation of \XD\  in $\Fun\X$.
Its irreducibility follows from the standard argument. Suppose
$V\subset \Fun\X$ is an invariant subspace. Being invariant, in particular,
under the commutative subalgebra generated by $\{x_n\}_{n=1}^N$ the space $V$
can consist only of the functions vanishing on some subset of $\X$.
However, such a subspace cannot be invariant under the operators $X_n^\pm$
because of the condition (\ref{eq:nondegL}) which ensures that the functions
$\D_n^\pm(\x)$ have no zeroes on $\X$ other than in $\X_n^\pm$.

\begin{th}
Let functions $\tilde{\D}_n^\pm$ define a finite-dimensional representation
of \XD\ with the same $\D(u)$ and $\X$ as for the above described standard
irreducible representation. Then the two representations are equivalent.
\label{th:equiv}
\end{th}

{\bf Proof.} We are going to show that the equivalence is provided by the
multiplication operator $\rho:\Fun\X\rightarrow\Fun\X:f(\x)\rightarrow
\rho(\x)f(\x)$ where $\rho(\x)$ is a function having no zeroes on $\X$.
The equivalence of operators $X_n^\pm$ leads to the equation
$$ \rho(E_n^\pm\x)\D_n^\pm(\x)=\rho(\x)\tilde{\D}_n^\pm(\x) $$
which can be considered as a set of recurrence relations for $\rho(\x)$.
Note that the recurrence relations are compatible because of the conditions
(\ref{eq:Dcond1}) for $\D_n^\pm$ and $\tilde{\D}_n^\pm$. The function
$\rho(\x)$ is thus defined uniquely on $\X$ up to an unsignificant coefficient.
It remains to show that $\rho(\x)\neq0$ on $\X$ which follows from the fact
that the functions $\tilde{\D}_n^\pm(\x)$ have no zeroes on $\X$ other than on
$\X_n^\pm$. The last assertion follows in turn from the third of the
equalities (\ref{eq:Dcond1}), definition (\ref{eq:defD}) of $\D(u)$
and the condition (\ref{eq:nondegL}).

It is natural to ask whether the above constructed sample representations
exhaust all the irreducible finite-dimensional representations of \XD.
I don't know the answer and hope that somebody of the audience will be
interested in this problem. Fortunately, the results already obtained
are enough to help solution of our main problem --- spectral analysis
of $\tau(u)$.

\vskip 1cm
\setcounter{chapter}{4}
\setcounter{equation}{0}
\setcounter{th}{0}
{\samepage
\begin{center}
    \large\bf Lecture 4
\end{center}

\nopagebreak
Let us apply now the results of the previous lecture concerning the
representation theory for the algebras \TR\ and \XD\ to our permanent
example --- XXX magnet.}

\begin{th} Let the representation $T(u)$ of \TR\ be given by the
product (\ref{eq:KLL})
$$ T(u)=KL_N(u)\ldots L_2(u)L_1(u)  $$
and the following nondegeneracy conditions be fulfilled.
The first one concerns the matrix K
$$ \det K\neq0 \qquad K_{12}\neq0 $$
and ensures the Condition 3.1 of Lecture 3. The second condition
coincides with the nonintersection condition (\ref{eq:nondegL}) for
the sets $\{\L_n\}_{n=1}^N$ defined by the formula
(\ref{eq:defLam}) with
$$ \l_n^\pm=\d_n\pm l_n  $$

Then all the conditions 3.1--4 of Lecture 3 are satisfied.
The spectrum of operators $\{x_n\}_{n=1}^N$ denoted now as $\tilde{\X}$
is the union of parallelepipeds
$$ \tilde{\X}=\bigcup_{\sigma\in S_N} \sigma\X $$
taken over all permutations $\sigma$ of coordinates $\{x_n\}_{n=1}^N$.
The set $\X$ is defined by the formula (\ref{eq:setX}). Note that the
sets $\sigma\X$ do not intersect due to condition (\ref{eq:nondegL}).
The corresponding representation of the algebra \XD\ is the direct
sum of $N!$ standard irreducible representations with
${\rm spec}\{x_n\}_{n=1}^N=\sigma\X$ described in the end of previous lecture,
the parameters $\l_n$ being defined above and $\xi_\pm$ being
arbitrary nonzero numbers satisfying
$$ \xi_+\xi_-=\det K $$
\label{th:specB}
\end{th}

The proof is performed by induction in $N$. Let $N=P+Q$ and
$$ T(u)=T_P(u)T_Q(u) $$
$$ \left(\begin{array}{cc}
           A&B\\C&D\end{array}\right)=
   \left(\begin{array}{cc}
           A_P&B_P\\C_P&D_P\end{array}\right)
   \left(\begin{array}{cc}
           A_Q&B_Q\\C_Q&D_Q\end{array}\right)    $$

Let the generators of \XD\ corresponding to
$ T,T_P,T_Q $ be, respectively,
$zZ^\pm$, $xX^\pm$, $yY^\pm$ and the corresponding functions $\D_\pm(u)$
be $\D_\pm(u)$, $\D_\pm^{(P)}$, $\D_\pm^{(Q)}$.

Consider the eigenvalue problem
$$ B(u)\Phi=\beta(u)\Phi  $$
for the matrix element $B(u)$ of $T(u)$
$$ B(u)=A_P(u)B_Q(u)+B_P(u)D_Q(u)  $$

Note that since the spaces $\Sym\Fun\tilde{\X}\otimes\Sym\Fun\tilde{\Y}$
and $\Fun\X\otimes\Fun\Y$ are isomorphic one can think of $\Phi$ as
belonging to $\Fun\X\otimes\Fun\Y\equiv\Fun(\X\times\Y)$.

Substitutions $u=x_p$\  $(p\in\{1,\ldots,P\})$ and
$u=y_q\ (q\in\{1,\ldots,Q\})$ result, respectively, in
$$\beta(x_p)\Phi(\x,\y)=b_Q(x_p-y_1)\ldots(x_p-y_Q)
            \D_-^{(P)}(x_p)\Phi(E_p^-\x,\y)   $$
and
$$\beta(y_q)\Phi(\x,\y)=b_P(y_q-x_1)\ldots(y_q-x_P)
            \D_+^{(Q)}(y_q)\Phi(\x,E_q^+\y)   $$
where
$$\D_-^{(P)}(x)=\xi_-^{(P)}\prod_{p=1}^P(x-\l_p^-) $$
$$\D_+^{(Q)}(y)=\xi_+^{(Q)}\prod_{q=1}^Q(y-\l_q^+) $$

Let us extract out of $\Phi$ the factor $\rho$
$$ \Phi(\x,\y)=\rho(\x,\y)\Psi(\x,\y)  $$
satisfying the equations
$$ b_Q(x_p-y_1)\ldots(x_p-y_Q)\rho(E_p^-\x,\y)=\rho(\x,\y) $$
$$ b_P(y_q-x_1)\ldots(y_q-x_P)\rho(\x,E_q^+\y)=\rho(\x,\y) $$
It is easy to verify that the equations are compatible and
have unique solution up to a constant coefficient.

The original spectral problem is written in terms of $\Psi$ as follows
$$ \beta(x_p)\Psi(\x,\y)=\D_-^{(P)}(x_p)\Psi(E_p^-\x,\y)  $$
$$ \beta(y_q)\Psi(\x,\y)=\D_+^{(Q)}(y_q)\Psi(\x,E_q^+\y)  $$
and apparently allows the \sov
$$\Psi(\x,\y)=\left(\prod_{p=1}^P\theta_p(x_p)\right)
              \left(\prod_{q=1}^Q\chi_q(y_q)\right)         $$
$$ \beta(x)\theta_p(x)=\D_-^{(P)}\theta_p(x-\eta)
    \qquad x\in\L_p^{(P)}                              $$
$$ \beta(y)\chi_q(y)=\D_+^{(Q)}\chi_q(y+\eta)
    \qquad y\in\L_q^{(Q)}                              $$

It remains to determine the spectrum of $\beta(u)$.
Consider the equation
$$ \beta(x)\theta(x)=\D_-^{(P)}(x)\theta(x-\eta)  $$

There is the alternative: either
$\theta(\l^-)\neq0$ and hence $\beta(\l^-)=0$, or $\theta(\l^-)=0$.
If $\theta(\l^-)=0$ then there is the new alternative: either
$\theta(\l^-+\eta)\neq0$ and then $\beta(\l^-+\eta)=0$, or
$\theta(\l^-+\eta)=0$, and so on.
This argument shows that there exists $\l\in\L$ such that $\beta(\l)=0$.
Consequentely,
$$ \beta(u)=\left(\prod_{p=1}^P(u-\l_p)\right)
            \left(\prod_{q=1}^Q(u-\l_q)\right)
   \qquad \l_p\in\L_p^{(P)}\quad\l_q\in\L_q^{(Q)} $$

or
$$ {\rm spec}\{z_n\}_{n=1}^N\equiv\tilde{\Z}=
         \bigcup_{\sigma\in S_N}\sigma \Z \qquad \Z=\X\times\Y $$
The spectrum of $z_n$ being the same as expected, it remains to refer
to the Theorem~\ref{th:equiv} claiming the uniqueness of the representations
of the algebra \XD\ corresponding to the sets $\sigma\Z$.

Let us return now to our main problem --- the spectral analysis
of $t(u)=A(u)+D(u)$. Consider the spectral problem
$$ t(u)\phi=\tau(u)\phi $$
then substitute from the left $u=x_n$ and use the definitions
(\ref{eq:defXm}), (\ref{eq:defXm})
of $X_n^\pm$ together with the expression (\ref{eq:exprX}) for $X_n^\pm$.
The resulting set of equations
$$ \tau(x_n)\phi(\x)=\D_n^+(\x)\phi(E_n^-\x)+\D_n^-(\x)\phi(E_n^+\x)
  \qquad n\in\{1,\ldots,N\} $$
allows, obviously, separation of variables
$$ \phi(x_1,\ldots,x_N)=\prod_{n=1}^N Q_n(x_n)$$
which leads to the set of N one-dimensional finite-difference
multiparameter spectral problems
\beq
\tau(x_n)Q_n(x_n)=\D_+(x_n)Q_n(x_n-\eta)+\D_-(x_n)Q_n(x_n+\eta)
   \quad x_n\in\L_n\quad n\in\{1,\ldots,N\}
\label{eq:sepeq}
\eeq
which are identical in the form with Baxters's equation
(\ref{eq:Baxter}) arising in ABA.
However, interpretation of the same equation is different. In the ABA case
$Q(u)$ was polynomial and now $Q$'s are functions on discrete sets $\L_n$.
In the ABA the natural way to solve the equation is to use the Bethe
equations (\ref{eq:Bethe}) for the zeroes of $Q(u)$.
Though Bethe equations as a rule cannot
be solved exactly they are simplified substantially in the infinite-volume
limit $N\rightarrow\infty$ which is the most interesting case for applications.
In contrast, the FBA result suggests solving the system of recurrence
relations for $Q(u)$ numerically which can be of interest for small $N$.
So, for the XXX magnet ABA and FBA approaches are complementary.
The FBA interpretation of the spectral problem (\ref{eq:sepeq})
has however some advantage:
since we have established the one-to-one correspondence between the
eigenfunctions $\phi$ of the original multidimensional spectral problem
and those of the related one-dimensional ones there is no {\it completeness
problem} in FBA which arises in ABA due to the fact that, generally
speaking, there could exist nonzero polynomial solutions to (\ref{eq:sepeq})
having no counterpart among eigenvectors of $t(u)$ and vice versa
there could be eigenvectors of $t(u)$ which cannot be expressed as
$B(v_1)\ldots B(v_M)\vac$. To be just, it is necessary to remark that
the incompleteness of ABA is a matter of degeneration and can always be removed
by a small variation of parameters.

Let us discuss now the possibility to apply FBA to the models other
than XXX magnetic chains. A possible way to generalize the above results
is to consider infinite-dimensional representations of the algebra
\TR. If the spectrum $\X$ of the operators $x_n$ is discrete then the
generalization presents no difficulty. Consider for example the
Goryachev-Chaplygin top discussed already in the previous lecture.
It turns out that in order to construct the matrix $T(u)$ representing
the algebra \TR\ it is necessary to introduce additional dynamical
canonical variables $p$ and $q$
$$ [p,q]=1 $$
commuting with the generators $x_\alpha$ and $J_\alpha$ of $e(3)$.
Then the elements of $T(u)$ are defined as follows
$$\begin{array}{rcl}
  A(u)&=&b(x_+u-\half\{x_3,J_+\})\\
  B(u)&=&\e^{-2\i q}(u^2-2J_3u-(J^2_3+\frac{1}{4}))\\
  C(u)&=&b\e^{2\i q}[(x_+u-\half\{x_3,J_+\})(u+p+2J_3)-bx_3^2]\\
  D(u)&=&(u+p+2J_3)(u^2-2J_3u-(J^2-J_3^2+\frac{1}{4})+bx_-u-
         \half\{x_3,J_-\})\end{array}$$
where $\{,\}$ is anticommutator, $x_\pm=x_1\pm\i x_2$,  $J_\pm=J_1\pm\i J_2$.
The corresponding R-matrix is $R(u)=u-2{\cal P}$.
It follows from the above formulas that
$$ t(u)=A(u)+D(u)=u^3+pu^2-2(H_p+\frac{1}{8})u-G_p$$
where
$$ H_p=H+pJ_3 \qquad G_p=G+p(J^2-J_3^2+\frac{1}{4})  $$
$H$ and $G$ being the integrals of motion of the quantum GC top
(\ref{eq:Htop}),
 (\ref{eq:Gtop}).
Since $p$ is one of integrals of motion it can be considered as a
scalar parameter in $H_p$ and $G_p$. In particular, for $p=0$ the
GC top is recovered. The general case $p\neq0$ has also nice physical
interpretation: it corresponds to the so-called gyrostat.
The zeroes of $u_1,u_2$ of $B(u)$ and the corresponding \sov\
are the same as discussed already in the previous lecture.

Much more difficult is the case when the zeroes $x_n$ of $B(u)$
have continuous spectrum. Such an example is provided by the
periodic Toda chain. In that case the spectrum of $x_n$ is
real and continuous but the shift $\eta$ in the Baxter's equations
is purely imaginary. The rigorous mathematical justification
of FBA presents in this situation serious analytical difficulties  which are
not yet overcome. However, quite formal application of FBA,
``on the physical level of rigour'', leads to reasonable results and
agrees with the results for the infinite volume limite
obtained by independent methods.

Another challenging problem is the generalization of FBA to
Yangians of simple Lie algebras other then $sl(2)$ that is
those of ${\rm rank}>1$. I believe that solution of this problem
will clarify the algebraic roots of FBA scheme which at present
is nothing that a misterious prescription: ``take zeroes of $B(u)$
and you obtain what you want''. Let me conclude my lectures
with a brief discussion of the problem for the $sl(3)$ case.

However, let us return first to the $sl(2)$ case and look once more
on the Baxter's equation (\ref{eq:Baxter}) dividing it by $Q(x)$
$$\tau(u)=\D_+(u)\frac{Q(u-\eta)}{Q(u)}+\D_-(u)\frac{Q(u+\eta)}{Q(u)} $$
The formula can be rewritten as
$$ \tau(u)=\L_+(u)+\L_-(u)\qquad
            \L_\pm(u)\equiv\D_\mp(u)\frac{Q(u\pm\eta)}{Q(u)}  $$
It is easy to see that
$$ \D(u)=\L_-(u+\halfe)\L_+(u-\halfe)$$
The fact that the quantities $\L_\pm(u)$ are expressed in terms of the
spectral invariants of the matrix $T(u)$, its trace $t(u)$ and quantum
determinant $\D(u)$,
suggests the interpretation of $\L_\pm(u)$ as the eigenvalues of
some operators which could be called ``the quantum eigenvalues of
the $2\times2$ matrix $T(u)$''. One can write down also the quantum
analogs of the secular equation for $\L_\pm(u)$
$$ \L_+(u)\L_+(u+\eta)-\tau(u+\eta)\L_+(u)+\D(u+\halfe)=0 $$
$$ \L_-(u-\eta)\L_-(u)-\tau(u-\eta)\L_-(u)+\D(u-\halfe)=0 $$
which turn into the familiar  quadratic equation
$$ \L^2-\tau\L+\D=0 $$
in the classical limit $\eta\rightarrow0$. Note that, in contrast
with the classical case, $\L_+$ and $\L_-$ satisfy two different
equations.

The operators $\X_n^\pm$ also can be considered as
``quantum eigenvalues'' of $T(x_n)$. Note that after subsitution
$u=x_n$ the matrix $T(u)$ becomes triangular due to $B(x_n)=0$.
In the classical case it means that the diagonal
elements  $X_n^-=A(x_n)$ and $X_n^+=D(x_n)$ of $T(u)$ coincide
with its eigenvalues. In the quantum case one can also write
down the quantum analogs of the secular equations for $X_n^\pm$.
They are obtained by excluding $X_n^+$ or $X_n^-$ from
the equalities
$$ t(x_n)=X_n^++X_n^- $$
$$ \D(x_n\pm\halfe)=X_n^\pm X_n^\mp $$
and read
$$ [X_n^+]^2-X_n^+t(x_n)+\D(x_n+\halfe)=0 $$
$$ [X_n^-]^2-X_n^-t(x_n)+\D(x_n-\halfe)=0 $$

The above observations need, of course, more profound study
aimed at understanding the algebraic meaning of the ``quantum eigenvalues''.
However, they have already enough heuristic power to make some
conclusions concerning FBA for $sl(3)$ case.

Consider the algebra \TR\ generated by the $3\times3$ matrix
$$ T(u)=\left(\begin{array}{ccc}
         T_{11}(u)&T_{12}(u)&T_{13}(u)\\
         T_{21}(u)&T_{22}(u)&T_{23}(u)\\
         T_{31}(u)&T_{32}(u)&T_{33}(u)
        \end{array}\right)                            $$
and the quadratic relations (\ref{eq:RTT}) with the R-matrix
$$ R(u)=u+\eta{\cal P}                                $$
in $\C^3\otimes\C^3$. The algebra is equivalent to the Yangian
${\cal Y}[sl(3)]$.

It turns out that the trace $t_1(u)$ of $T(u)$
$$ t_1(u)=\tr T(u)$$
does not provide the complete set of integrals of motion
and should be accompanied with the operators
$$t_2(u)=\tr_{12} P^-_{12}\one T(u)\two T(u+\eta)  $$
where $P^-_{12}=(1-{\cal P})/2$ is the antisymmetrizer in $\C^3\otimes\C^3$,
see (\ref{eq:qdet}).

The Casimir operator of \TR\ (quantum determinant of $T(u)$) is
given by the formula
$$   \D(u)=\qdet T(u)=
  \tr_{123}P_{123}^-\one T(u)\two T(u+\eta)\three T(u+2\eta)  $$
where $P^-_{123}$ is the antisymmetrizer in $\C^3\otimes\C^3\otimes\C^3$.

The commuting quantities $t_1(u)$, $t_2(u)$ and $\D(u)$ constitute three
spectral invariants of the matrix $T(u)$.

The ABA for $sl(N)$ case was developed in the papers by Yang, Sutherland,
Kulish and Reshetikhin. It is  the results of the last two
authors which will be especially useful for our purposes. As Kulish
and Reshetikhin (1982) have shown, the eigenvalues $\tau_{1,2}(u)$ of
$t_{1,2}(u)$ together with the quantum determinant
$\D(u)$ can be written down in the form

\begin{eqnarray}
\tau_1(u)&=&\L_1(u)+\L_2(u)+\L_3(u)  \nonumber \\
\tau_2(u)&=&\L_1(u)\L_2(u+\eta)+\L_1(u)\L_3(u+\eta)+\L_2(u)\L_3(u+\eta)
\label{eq:eigv} \\
\D(u)&=&\L_1(u)\L_2(u+\eta)\L_3(u+2\eta)
       =\D_1(u)\D_2(u+\eta)\D_3(u+2\eta)  \nonumber
\end{eqnarray}
where the number polynomials $\D_{1,2,3}(u)$ like $\D_\pm(u)$ for
$sl(2)$ case are expressed in terms of representation parameters.

The three ``quantum eigenvalues'' $\L_{1,2,3}(u)$ of $T(u)$
can be expressed in terms of two polynomials $Q_{1,2}(u)$
$$\L_1(u)=\D_1(u)\frac{Q_1(u+\eta)}{Q_1(u)}\qquad
  \L_2(u)=\D_2(u)\frac{Q_1(u-\eta)}{Q_1(u)}\frac{Q_2(u+\eta)}{Q_2(u)}$$
$$  \L_3(u)=\D_3(u)\frac{Q_2(u-\eta)}{Q_2(u)} $$

Kulish and Reshetikhin have derived also from (\ref{eq:eigv})
the finite-difference
equations (analog of Baxter's equation) for $Q_{1,2}(u)$ which with the
use of the shift/mul\-tip\-li\-ca\-tion operators
$$\Xi_1=\D_1(x)\e^{\eta\partial/\partial x}  $$
$$\Xi_2=\D_1(x-\eta)\D_2(x)\e^{\eta\partial/\partial x}  $$
can be put into the form
\beq
\begin{array}{rcl}
 {[}\Xi_1^3-\tau_1(x+2\eta)\Xi_1^2+\tau_2(x+\eta)\Xi_1-\D(x){]}Q_1(x)&=&0 \\
 {[}\Xi_2^3-\tau_2(x+\eta)\Xi_2^2+\tau_1(x+\eta)\D(x)\Xi_2-
     \D(x-\eta)\D(x){]}Q_2(x)&=&0 \\
\end{array}
\label{eq:sepeq3}
\eeq
The last equations resemble very much the FBA separated equations
and suggest, by analogy with the $sl(2)$ case, the following conjecture
concerning the possible form of FBA in the $sl(3)$ case.

{\bf Conjecture.} {\it There exist two sets of operators
$x_1^{(n_1)},X_1^{(n_1)}\quad n_1\in\{1,\ldots,N\} $
and
$x_2^{(n_2)},X_2^{(n_2)}\quad n_2\in\{1,\ldots,2N\} $
such that $X_{1,2}$ satisfy the ``secular equations''
$$X_1^3-X_1^2t_1(x_1)+X_1t_2(x_1)-\D(x_1)=0  \qquad\forall n_1$$
$$X_2^3-X_2^2t_2(x_2-\eta)+X_2\D(x_2-\eta)t_1(x_2)-\D(x_2-\eta)\D(x_2)=0
                                \qquad\forall n_2$$
which lead to the separated equations (\ref{eq:sepeq3})
for the eigenvalue problem
for the operators $t_{1,2}(u)$ and the eigenfunction}
$$ \phi(\x_1,\x_2)=\prod_{n_1=1}^NQ_1(x_1^{(n_1)})
                   \prod_{n_2=1}^{2N}Q_2(x_2^{(n_2)})  $$

The problem which remains unsolved is how to construct such operators
that is to find a generalisation of the recipe $B(u)=0$
valid for the $sl(2)$ case. In my opininon, solution of this
problem will contribute to better understanding of the algebraic
roots of Bethe Ansatz.

\vskip 1.cm
\begin{center}
 \large\bf Bibliographical Notes
\end{center}

\noindent{\bf Lecture 1}

The main ideas of the algebraic approach to the quantum integrability
are present already in the papers of fathers of quantum mechanics
\bds
\item[{}] W.\ Heisenberg (1925) {\it Z.\ Phys.\ }{\bf 33}, 879.
\item[{}] W.\ Pauli (1926) {\it Z.\ Phys.\ }{\bf 36}, 336--363.
\eds
An extensive survey of the Classical Inverse Scattering Method with special
stress on its Hamiltonian aspects wich are important for quantization
can be found in the book
\bds
\item[{}] L.\ D.\ Faddeev, L.\ A.\ Takhtajan (1987) Hamiltonian Methods in the
  Theory of Solitons. Berlin: Sprin\-ger.
\eds
Reviews of the Quantum Inverse Scattering Method
\bds
\item[{}] L.\ A.\ Takhtajan, L.\ D.\ Faddeev (1979)
  {\it Russian Math.\ Surveys\ }{\bf 34}:5, 11--68.
\item[{}] P.\ P.\ Kulish, E.\ K.\ Sklyanin (1982a)
{\it Lecture Notes in Physics}    {\bf 151}, pp.61--119, Berlin: Sprin\-ger.
\item[{}] L.\ D.\ Faddeev (1984) Les Houches, Session XXXIX, 1982, Recent
Adv.\  in Field
 Theory and Stat.\  Mech., eds.\  J.-B.\ Zuber, R.\  Stora, pp.563--608,
Elsevier.
\item[{}] L.\ D.\ Faddeev (1987) in Nankai Lectures on Mathematical Physics,
Integrable Systems, ed. by X.-C.\ Song, pp.23--70, Singapore: World Scientific,
1990.
\eds
Elementary facts concerning the Yang-Baxter equation are contained in the
surveys
\bds
\item[{}] P.\ P.\ Kulish, E.\ K.\ Sklyanin (1982b) {\it J.\ Sov.\ Math.\ }
 {\bf 19}, 1596--1620.
\item[{}] M.\ Jimbo (1989) {\it Int.\ J.\ Mod.\ Phys.\ }{\bf 4}, 3759--3777.
\eds
For more sophisticated treatment of the Yang-Baxter equation based on the
Quantum Group theory see
\bds
\item[{}] V.\ G.\ Drinfeld (1985) {\it Sov.\ Math.\ Dokl.\ }{\bf 32}, 254--258.
\item[{}] V.\ G.\ Drinfeld (1988) {\it Sov.\ Math.\ Dokl.\ }{\bf 36}, 212--216.
\item[{}] V.\ G.\ Drinfeld (1987) Quantum Groups, in
   {\it Proceedings of the International Congress of Mathematicians},
Berkeley, 1986, American Mathematical Society, 798--820.
\item[{}] L.\ A.\ Takhtajan (1989) in Nankai Lectures on Mathematical Physics,
 Introduction
 to Quantum Group and Integrable Massive Models of Quantum Field Theory,
eds.\  M.-L.\ Ge, B.-H.\ Zhao, pp.69--197, Singapore: World Scientific, 1990.
\eds

\noindent{\bf Lecture 2}

The proof of the theorem~\ref{th:LL} follows the paper
\bds
\item[{}] N.\ Yu.\ Reshetikhin, P.\ P.\ Kulish, E.\ K.\ Sklyanin (1981)
  {\it Lett.\ Math.\ Phys.\ }{\bf 5}, 393--403.
\eds
The theorem~\ref{th:Sutherland} is borrowed from
\bds
\item[{}] B.\ Sutherland (1970) {\it J.\ Math.\ Phys.\ }{\bf 11}, 3183--3186.
\eds
The construction of the local integrals of motion using the generating
function $\tau(u)$, see formula (\ref{eq:deftau}), was proposed in
\bds
\item[{}] R.\ J.\ Baxter (1972) {\it Ann.\ Phys.\ }{\bf 70}, 323--337.
\item[{}] M.\ L\"{u}scher (1976) {\it Nucl.\ Phys.\ }B{\bf 117}, 475--492.
\eds
The definition of $t^{(k)}(u)$ is taken from (Kulish, Sklyanin, 1982a),
see above. For the master symmetries for quantum integrable chains see
\bds
\item[{}] M.\ G.\ Tetelman (1982) {\it Sov.\ Phys.\  JETP\ }{\bf 55}(2),
306--310.
\item[{}] E.\ Barouch, B.\ Fuchssteiner (1985) {\it Stud.\ Appl.\ Math.\ }
  {\bf 73}, 221--237.
\item[{}] H.\ Araki (1990) {\it Commun.\ Math.\ Phys.\ }{\bf 132},
  155--176.
\eds
The boost operator $\BB$ was shown recently to have close relation to
Baxter's corner transfer matrices
\bds
\item[{}] H.\ Itoyama, H.\ B.\ Thacker (1987) {\it Phys.\ Rev.\ Lett.\ }
{\bf 58},   1395--1398.
\eds

\noindent{\bf Lecture 3}

The Baxter's equation (\ref{eq:Baxter}) was introduced in
\bds
\item[{}] R.\ J.\ Baxter (1971) {\it Stud.\ in Appl.\ Math.\ }{\bf 50}, 51--69.
\eds
An alternative derivation of Baxter's equation using ``Analytical Bethe
Ansatz'' is given in
\bds
\item[{}] N.\ Yu.\ Reshetikhin (1983) {\it Lett.\ Math.\ Phys.\ }
{\bf 7}, 205--213.
\eds
For the Algebraic Bethe Ansatz see reviews on QISM cited above. Examples
of incompleteness of ABA are presented in
\bds
\item[{}] L.\ V.\ Avdeev, A.\ A.\ Vladimirov (1986)
  {\it Theor.\ {\rm\&}\ Math.\ Phys.\ }{\bf 69}, 1071--1079.
\item[{}] T.\ Koma, H.\ Ezawa (1987) {\it Progr.\ Theor.\ Phys.\ }{\bf 78},
   1009--1021.
\eds
Separation of variables in the quantum Goryachev-Chaplygin top (gyrostat)
was obtained in
\bds
\item[{}] I.\ V.\ Komarov (1982) {\it Theor.\ {\rm\&}\ Math.\ Phys.\ }{\bf 50},
 265--270.
\item[{}] I.\ V.\ Komarov, V.\ V.\ Zalipaev (1984) {\it J.\ Phys.\ A: Math.\
Gen.\ }{\bf 17}, 31--49.
\eds
The Functional Bethe Ansatz was proposed in papers
\bds
  \item[{}] E.\ K.\ Sklyanin (1985a) {\it Lecture Notes in Physics\ }{\bf 226},
pp.196--233, Berlin: Sprin\-ger.
\item[{}] E.\ K.\ Sklyanin (1985b) {\it J.\ Sov.\ Math.\ }{\bf 31}, 3417--3431.
\eds
influenced deeply by the articles
\bds
\item[{}] H.\ Flashka, D.\ W.\ McLaughlin (1976) {\it Progr.\ Theor.\
Phys.\ }{\bf 55}, 438--456.
\item[{}] M.\ Gutzwiller (1981) {\it Ann.\ Phys.\ }{\bf 133} 304--331.
\eds
The exposition of FBA in the present lectures is an improved and
extended variant of the paper
\bds
\item[{}] E.\ K.\ Sklyanin (1990) ``Functional Bethe Ansatz'',
in {\it Integrable and Superintegrable Systems},
ed.\  by B.\ A.\ Kupershmidt, pp.8--33, Singapore: World Scientific.
\eds
The proof of Theorem~\ref{th:XX} is given in full length.
The idea of extended space $\tilde{W}$ as well as the
Theorems~\ref{th:irred} and~\ref{th:equiv} are new.

\begin{flushleft}
{\bf Lecture 4}
\end{flushleft}

The proof of the Theorem~\ref{th:specB} sketched in (Sklyanin, 1990)
is given here in full length. For the application of FBA to
Goryachev-Chaplygin top and Toda chain see, respectively
Sklyanin (1985b) and (1985a). The analysis of SL(3) case is based on
the paper
\bds
\item[{}]  P.\ P.\ Kulish, N.\ Yu.\ Reshetikhin (1982) {\it Zapiski
 Nauch.\ Semin.\ Leningr.\ Otd.\ Matem.\ Inst.\ Steklova\ }{\bf 120}, 92--121.
\eds

\end{document}